\documentclass[12pt,preprint]{aastex}

\shorttitle{late formation of extrasolar giant planets} 

\shortauthors{Ida and Lin}

\begin{document}

\title{Toward a Deterministic Model of Planetary Formation IV:}
\title{Effects of Type-I Migration}

\author{S. Ida}
\affil{Tokyo Institute of Technology,
Ookayama, Meguro-ku, Tokyo 152-8551, Japan}
\email{ida@geo.titech.ac.jp}

\and 

\author{D. N. C. Lin}
\affil{UCO/Lick Observatory, University of California, 
Santa Cruz, CA 95064, USA}
\affil{Kavli Institute of Astronomy \& Astrophysics, Peking
University, Beijing, China}
\email{lin@ucolick.org}

\begin{abstract}
In a further development of a deterministic planet-formation model
(Ida \& Lin 2004), we consider the effect of type-I migration of
protoplanetary embryos due to their tidal interaction with their
nascent disks.  During the early embedded phase of protostellar disks,
although embryos rapidly emerge in regions interior to the ice line,
uninhibited type-I migration leads to their efficient self-clearing. 
But, embryos continue to form from residual planetesimals at
increasingly large radii, repeatedly migrate inward, and provide a
main channel of heavy element accretion onto their host stars. During
the advanced stages of disk evolution (a few Myr), the gas surface
density declines to values comparable to or smaller than that of the
minimum mass nebula (MMSN) model and type-I migration is no longer an
effective disruption mechanism for mars-mass embryos. Over wide
ranges of initial disk surface densities and type-I migration
efficiency, the surviving population of embryos interior to the ice
line has a total mass several times that of the Earth.  With this
reservoir, there is an adequate inventory of residual embryos to
subsequently assemble into rocky planets similar to those around the
Sun.  But, the onset of efficient gas accretion requires the emergence
and retention of cores, more massive than a few $M_\oplus$, prior to
the severe depletion of the disk gas.  The formation probability of
gas giant planets and hence the predicted mass and semimajor axis
distributions of extrasolar gas giants are sensitively determined by
the strength of type-I migration.  We suggest that the observed
fraction of solar-type stars with gas giant planets can be reproduced
only if the actual type-I migration time scale is an order of
magnitude longer than that deduced from linear theories.  We also show
that the introduction of such slower type-I migration rate makes
simulated planetary mass-period distribution more consistent with
observation.

\end{abstract}

\section{Introduction}
In the radial velocity survey of nearby stars, more than 200 
extrasolar planets with mass $M_p$ comparable to that of Jupiter $M_J$ have
been discovered. Current statistics indicates that $\eta_J>7\%$ of all
nearby F, G, and K dwarfs on various search target lists have gas
giant planets with semimajor axis $<5$AU \citep[e.g.,][]{Marcy05,
Mayor05}.  But the extrapolation of existing data suggests a higher
fraction of solar-type stars may have longer-period gas giant 
planets \citep{Cumming}.  

In the previous papers of this series \citep[][hereafter referred to
as papers I, II, and III] {IL04a, IL04b, IL05}, we carried out several
sets of simulations with a numerical prescription, which is based on
the sequential accretion scenario.  We used these results to explain
the observed mass and orbital distributions of extrasolar planets and
to constrain the intrinsic physics of planet formation through
detailed comparisons with the observed data.

In these previous investigation, we have taken into account the effect
of gas depletion by assuming that the gas surface density $\Sigma_g$
declines everywhere exponentially over a characteristic timescale
$\tau_{\rm dep} \simeq 1$--10 Myr. With an $\alpha$ prescription for
the turbulent viscosity, this assumed global evolution of $\Sigma_g$
can lead to declining accretion rates which are consistent with those
observed \citep{Hartmann98}. The magnitude of $\tau_{\rm dep}$ sets a
strong constraint on the gas giant planet building efficiency
\citep[e.g.,][]{IL04a}.

We have also assumed that the dust surface density ($\Sigma_d$) in
these disks is preserved from the initial value we have adopted.  The
dust grains' mass inferred from the mm observation of continuum
radiation from protostellar disks \citep[e.g.,][]{BS96} appears to
have a wide dispersion centered around the value of the minimum-mass
solar nebula (MMSN) model \citep{Hayashi81}.  We adopted a similar
distribution for $\Sigma_d$ centered around that of MMSN.  This
assumption, though it greatly simplified our treatment, is less well
justified.  In the scaling of $\Sigma_d$ with that of MMSN, we have
inherited the assumption that all the heavy elements were locally
retained during the epoch of planet formation. 

There are two potential avenues for the depletion of heavy elements:
hydrodynamic gas drag of small dust grains and type-I migration of
cores due to the tidal interaction with their nascent disks.  Here,
"cores" mean the protoplanetary embryos which formed as a result of
runaway/oligarchic growth of accretion of planetesimals \citep{KI98,KI02}.
These cores are generally not sufficiently massive to initiate runaway
gas accretion.  Without much additional growth before the disk gas
is severely depleted, these cores would become either rocky planets or
ice giants.  

In the present paper, we consider the dominant processes after the
formation of planetary embryos which are hydrodynamically decoupled
from disk gas motions.  Prior to the phase of rapid gas accretion,
these embedded cores and their surrounding gas exchange angular
momentum as they engage in tidal interaction \citep{GT80,
LP79}. Analytic studies suggest that isolated cores lose angular
momentum to the disk exterior to their orbits faster than that they
gain from the disk interior to their orbits. This torque imbalance
leads to a ``type-I'' migration. From the linear analysis, the
characteristic orbital decay time scale of Earth-mass cores at several
AU's in a MMNM is much shorter than 1 Myr \citep{Ward86, Tanaka02}.
Accordingly, embryos formed well outside the ice line may migrate to
the present locations of the Earth and Venus. But, the mostly
refractory compositions of the terrestrial planets in the present-day
solar system suggest that they probably did not migrate to their
present location from regions well beyond the ice line.  

Today, there remains considerable uncertainties in the efficiency of
this process (see \S 2.2).  Nevertheless, their formation could have
been preceded by a much larger population of cores which did migrate
into the Sun.  In view of this uncertainty, we carry out a parametric
study on the decline of the cores' accretion rate associated with a
$\Sigma_d$ reduction due to the type-I migration of cores (see below).
This effect has been neglected in our previous papers.

Heavy elements are also depleted from the disks through ``type-II''
migration.  Due to runaway gas accretion onto a core, the growing gas
giant planet eventually acquires sufficient mass to open a gap in the
disk and lock itself into a ``type-II'' migration with the viscous
evolution of the disk gas \citep{LP85,LP93}.  In the context of solar
system formation, \citet{Lin86, Lin95} has speculated that, when the
solar nebula was massive, several gas giants may have formed,
undergone type-II migration, and eventually merged with the Sun.  When
the disk's $\Sigma_g$ decreases below that of the MMSN model, the
migration has stalled.  According to this scenario, Jupiter and Saturn
could be the last survivors of several generations of protoplanets
\citep[also see][]{Papaloizou06}, although it may not be easy to form
terrestrial planets after preceding gas giants have passed through the
inner disk region \citep{Armitage03}.  A resurgence of this migration
and survival scenario \citep{Lin96} have been stimulated by the
observational discoveries of close-in gas giant planets \citep[e.g.,]
[]{Marcy05,Mayor05}. But, a majority of the known extrasolar planets
have periods much longer than a few days.  We showed in Papers I and
II that their observed logarithmic period distribution can be used to
infer comparable timescales for giant planets' migration and disk depletion.
However, based on the relative abundance between planets with $a<0.06$
AU and those between 0.2-2AU's from their host stars, we have to
assume that up to 90\% of the gas giants which migrated to the
proximity of their host stars may have perished.

Here, we focus on the repeated clearing due to type-I migration.  
We suggest that this self-regulated clearing mechanism limits and
determines the amount of heavy elements retained by the cores and
residual planetesimals (also see discussion on metallicity homogeneity
of open clusters in \S 4.1).  It also results in late formation of
gas giants and the marginal probability of gas giants' formation
through a series of failed attempts.

The scenario we consider here is similar to the hypothesis proposed by
\citet{CW06} in the context of satellite formation around Jovian
planets. They suggested that the total mass of the surviving
satellites is self-regulated by their type-I migration through their
nascent circum-planetary disk which is continuously replenished.  In
the context of planet formation, \citet{McNeil05} carried out N-body
simulations of terrestrial planets' accretion and growth including the
effect of type-I migration. They found that it is possible to retain
sufficient amount of solid materials (planetesimals and embryos) to
assemble Earth-mass planets within a disk lifetime $\sim 10^6$ years
provided the type-I migration rate is slightly slower than that
predicted by the linear theory.  In these simulations, the initial
disk mass in terrestrial planet regions is assumed to be 3--4 times of
that of MMSN.  Based on an analytic approximation for type-I
migration, \citet{Kominami06} simulated the evolution of $\Sigma_d$
under various disk conditions.  Their results indicate that the
depletion region expands outwards, starting from the disk's inner
boundary.  They also found that in disks with same dust-to-gas ratio,
the asymptotic surface density of the retained embryos decreases with
the initial value of $\Sigma_d$ (and $\Sigma_g$) because type-I
migration is faster in more massive disks.

Here we consider a much large range of initial conditions. With a
comprehensive numerical prescription, we consider the possibility of
multiple generation of embryo formation. We show that in relatively
massive disks although type-I migration is more effective for
individual planetesimals and embryos, it does not necessarily lead to
more rapid depletion of the residual population.  In these disks,
migration starts with smaller individual masses but more generations
of embryo form and parish before the initial supply of planet-building
blocks is severely depleted. Our results indicate that the total
retainable mass of heavy elements only weakly depends on the initial
disk mass.

Although type-I migration places a mass limit on the retainable
embryos, it does not quench the formation of Earth-like terrestrial
planets because they can be assembled from retainable low-mass embryos
after the gas is depleted \citep{Kominami06}. Gas giant planets,
however, must form in gas-rich disks which appears to be depleted on the
time scale $\tau_{\rm dep} \sim$ a few Myr.  In addition, they must
acquire $\sim 10M_{\oplus}$ cores prior to the onset of efficient gas
accretion.  Although they can form rapidly outside the ice lines on
massive disks, early generation of such massive cores quickly migrate
into their host stars.  But, \citet{Thommes06} showed that cores
formed after the disk gas has been severely depleted may withstand the
disruption by the declining type-I migration. In \S3 of this paper, we
present results which are consistent with that obtained by
\citet{Thommes06}. We also incorporate gas accretion onto cores and
the effect of type-II migration and find that under some
circumstances, there is adequate residual gas to promote the efficient
gas accretion and the formation of gas giant planets.  This ``late
formation'' scenario is conceptually consistent with that inferred
from the noble gas enrichment in Jupiter \citep{Guillot06}.  Based on
this model, we are able to reproduce the observed mass-period
distribution of the known gas giant planets around solar type stars
provided the efficiency of type-I migration is at least an order of
magnitude slower than that derived from the linear theory (see \S3).
Our conclusion is consistent with the studies by 
\citet{libert05} who
also found that a necessary condition for the formation of Jupiter and
Saturn is a 30-times reduction in the type-I migration rate.

In metal-poor disks, the formation of critical mass cores requires a
relatively large gas surface density.  In these disks, type-I
migration is more effective in clearing the cores prior to the onset
of gas accretion.  Consequently, the formation of gas giant planets is
suppressed.  Assuming the disks' initial metallicity is identical to
that of their host stars, this effect sharpens the dependence of
the predicted formation efficiency of gas giants on the stellar
metallicity which is well established in the observational data
\citep{Fischer05}.

Since type-I and II migrations are the essential processes which
determine the properties of emerging planets, we briefly recapitulate,
in \S2, their basic physical principles.  We also incorporate a
quantitative prescription of these processes in our existing
comprehensive model for planet formation.  The model for the new
addition, type-I migration and the decrease in planetesimal surface 
density due to the migration, is explained in
detail.  In \S3, we carry out a systematic study on how cores'
migration may affect the formation of terrestrial planets and gas
giant planets.  We show that type-I migration delays formation of gas
giants, and that it leads to the mass and semimajor axis ($M_p-a$)
distribution consistent with that of the known gas giants around G
dwarfs.  We also study dependence of the the distribution on planet
formation parameters, metallicity.  The introduction of relatively
slow type-I migration results in the metallicity dependence that may
be consistent with observation.  Finally, we summarize our results and
discuss their implications in \S4.

\section{Planet formation and migration model}
\label{sec:prescription}
In this paper, we simulate the $M_p-a$ distribution of extrasolar
planets, taking into account the effects of cores' type-I migration as
well as type-II migration.  In Papers I-III, we outlined in detail a
quantitative prescription which we used to model the evolution of
planetesimals, gas accretion onto protoplanetary cores, the
termination of gas giant planet growth, and type-II migration.  In
this section we briefly recapitulate the dependence of both types of
migration on the background disk properties.

In this paper, we newly incorporate the effects of type-I migration
and examine the impact of type-I migration on the formation of
terrestrial planets and the asymptotic properties of emerging gas
giant planets.  In light of the theoretical uncertainties, we
introduce a prescription which captures the main determining factors
of type-I migration and enables us to consider a wide range in the
magnitude of its efficiency factor.
We here consider only the case of $M_* = 1M_{\odot}$ to 
focus on the effects of type-I migration around solar-type stars,
although we retain the dependences on stellar mass $M_*$ and 
its luminosity $L_*$.

Note that inclusion of type-I migration and the minor changes 
in truncation conditions of growth of gas giants and the disk model 
that are described below do not change 
the conclusions that have been derived in previous papers:
(i) the existence of "planet desert" (a lack of intermediate mass planets 
at $\la$ a few AU) in Paper I, (ii) the metallicity dependence 
(a fraction of stars with giant planets increases with the metallicity) 
in Paper II, and (iii) the stellar mass dependence 
(giant planets are much less abundant around lower-mass stars) in Paper III.
The results with inclusion of type-I migration corresponding to (i) and (ii) are shown 
in Figures \ref{fig:ma_C1} and \ref{fig:metal_dep}, respectively.
The stellar mass dependence with type-I migration
is briefly commented on in \S 3.2.3 and 
will be discussed in detail in a separate paper.

\subsection{Parameterized disk models}

The main objectives of this series of papers is to examine the
statistical properties of emerging planets under a variety of disk
environments rather than to study individual processes which regulate
the disk structure.  For computational convenience, we introduced, in
Papers I-III, multiplicative factors ($f_d$ and $f_g$) to scale
$\Sigma_g$ and $\Sigma_d$ such that
\begin{equation}
\left\{ \begin{array}{ll} 
\Sigma_d & = \Sigma_{d,10} \eta_{\rm ice} 
f_d (r/ {\rm 10 AU})^{-q_d}, 
\label{eq:sigma_dust} \\ 
\Sigma_g & = \Sigma_{g,10} f_{g} (r/ {\rm 10 AU})^{-q_g},
\label{eq:sigma_gas}
\end{array} \right.
\end{equation}
where normalization factors $\Sigma_{d,10} = 0.32 {\rm g/cm}^2$ and
$\Sigma_{g,10} = 75 {\rm g/cm}^2$ correspond to 1.4 times of
$\Sigma_g$ and $\Sigma_d$ at 10AU of the MMSN model, and the step
function $\eta_{\rm ice} = 1$ inside the ice line at $a_{\rm ice}$
(eq.~[\ref{eq:a_ice}]) and 4.2 for $r > a_{\rm ice}$ [the latter can
be slightly smaller ($\sim 3.0$) \citep{Pollack94}].

We show below that the disk metallicity [Fe/H]$_d$ is an important
parameter which regulates the survival of protoplanetary cores during
their type-I migration.  Dependence of disk metallicity is attributed
to distribution of $f_{d,0} = f_{g,0} 10^{{\rm [Fe/H]}_d}$.  Solar
metallicity corresponds to [Fe/H]$_d = 0$ and $f_{d,0} = f_{g,0}$.
The main advantage of these parameterized disk structure models is
that they enable us to efficiently simulate and to examine the
dominant dependence of planet formation and dynamical evolution on the
disk structure.

In self-consistent treatment of accretion flow, the disk temperature
is determined by an equilibrium between the viscous dissipation and
heat transport \citep{alpha}.  We neglect the detailed energy balance
\citep{Chiang97, Garaud07} and adopt the equilibrium temperature
distribution in highly optically thin disks prescribed by
\citet{Hayashi81} such that
\begin{equation}
T = 280 \left(\frac{r}{1{\rm AU}}\right)^{-1/2}
    \left(\frac{L_*}{L_{\odot}}\right)^{1/4} {\rm K}.
\label{eq:temp_dist}
\end{equation}
In this simple prescription, we set the ice line to be that determined
by an equilibrium temperature (eq.~[\ref{eq:temp_dist}]) in optically
thin disk regions \citep{Hayashi81},
\begin{equation}
a_{\rm ice} = 2.7 (L_\ast/L_\odot)^{1/2} {\rm AU},
\label{eq:a_ice}
\end{equation}
where $L_*$ and $L_{\odot}$ are the stellar and solar luminosity.  The
magnitude of $a_{\rm ice}$ may be modified by the local viscous
dissipation \citep{Lecar06} and stellar irradiation \citep{Chiang97,
Garaud07}.  These effects do not greatly modify the disk structure
during the late evolutionary stages and they will be incorporated in
subsequent papers.

\subsubsection{Evolution of $f_g$}

The surface density of the gas can be determined by a diffusion
equation \citep{LP85},
\begin{equation}
\frac{\partial \Sigma_g}{\partial t} - {1 \over r}
\frac{\partial}{\partial r} \left[ 3 r^{1/2}
\frac{\partial}{\partial r} (\Sigma_g \nu r^{1/2}) \right] = 0.
\label{eq:surf_density_evol}
\end{equation}
where we neglected the sink terms due to photoevaporation and
accretion onto the cores.  For computational convenience, we adopt the
standard constant-$\alpha$ prescription for viscosity \citep{alpha},
\begin{equation}
\nu= \alpha c_s H,
\label{eq:alpha}
\end{equation}
where $c_s$, $H=\sqrt{2}c_s/\Omega_{\rm K}$, and $\Omega_{\rm K}$ are
the sound speed, disk scale height, and the Keplerian angular frequency.
Although the value of $\alpha$ is not clear, $\alpha \sim 10^{-3}$ may
be consistent with the observational data of accretion rates of T
Tauri disks \citep[e.g.,][]{Hartmann98} and the results of MHD
simulations \citep[e.g.,][]{Sano04}.  We here use $\alpha = 10^{-3}$
as a nominal value.

With the $\alpha$ prescription and eq.~(\ref{eq:temp_dist}), $\nu
\propto r$ and eq.~(\ref{eq:surf_density_evol}) reduced to a linear
partial differential equation for which self-similarity solutions have
been presented by \citet{Lynden-Bell74}. The numerical solution for
disk-evolution equation [{\it i.e.}, eq.~(\ref{eq:surf_density_evol})
with $\alpha = 10^{-3}$], 
starting with $\Sigma_g \propto r^{-1.5}$ for
$r < r_m$ and an exponential cutoff at $r_m = 10$AU, is illustrated in
Figure \ref{fig:disk_vis_evol}a.

After a brief initial transition, the numerical solution quickly
approaches to $\Sigma_g \propto r^{-1}$ with an asymptotic exponential
cut-off, which is the self-similar solution obtained by
\citet{Lynden-Bell74} for a linear viscosity prescription and by
\citet{Linbod82} for an $\alpha$ model. As shown in
Figure~\ref{fig:disk_vis_evol}b, $\Sigma_g$ in the region $r < r_m$
decreases as uniformly independent of $r$.  In the self-similar
solution, $\Sigma_g$ at $r < r_m$ decays as $\Sigma_g \propto
(t/\tau_{\rm dep} + 1)^{-3/2}$, where
\begin{equation}
\tau_{\rm dep} = \frac{r_m^2}{3\nu(r_m)} 
\simeq 3 \times 10^6 
    \left(\frac{\alpha}{10^{-3}}\right)^{-1}
    \left(\frac{r_m}{100{\rm AU}}\right)
    \left(\frac{M_*}{M_{\odot}}\right)^{-1/2}
    {\rm yrs}.
\end{equation}
In the above, we used eqs.~(\ref{eq:temp_dist}) and
(\ref{eq:alpha}).
 
This self-similar nature of the solution is preserved if we assume a
spatially uniform exponential depletion of the disk ({\it i.e.}
$\Sigma_g \propto \exp(-t/\tau_{\rm dep})$ with a $r$-independent
$\tau_{\rm dep}$), as in Papers I-III, although in the actual
self-similar solution, $\Sigma_g \propto t^{-3/2}$ for $t > t_{\rm
dep}$.  If the effect of photoevaporation is taken into account,
$\Sigma_g$ decays rapidly after it is significantly depleted, so that
the exponential decay could be more appropriate.  Thus, we adopt
\begin{equation}
f_{g} = f_{g,0} \exp(-t/\tau_{\rm dep}),
\label{eq:gas_exp_decay}
\end{equation}
in the disk model in eq.~(\ref{eq:sigma_gas}).  Even if we fix the
$\alpha$ value, $\tau_{\rm dep}$ has uncertainty, because we do not
have enough knowledge about $r_m$, and $r_m$ should have some
dispersion.  We use $\tau_{\rm dep}$ as a parameter and set it to be
in a range of $10^6$--$10^7$ yrs, which may be consistent
with observation.
Corresponding to the self-similar solution, we consider here disk
models with $\Sigma_g \propto r^{-1}$ ($q_g = 1$) in addition to the
case with $\Sigma_g \propto r^{-3/2}$ ($q_g = 3/2$) that was assumed
in Papers I-III corresponding to the MMSN model.

\subsubsection{Evolution of $f_d$}

We start each model with $\Sigma_d \propto r^{-3/2}$ ($q_d = 3/2$).
In principle, the evolution of $\Sigma_d$ should be treated
independently from $\Sigma_g$ even though small dust particles are
thermally and dynamically coupled to the gas.  Physically, the
magnitude of $\Sigma_d$ is determined by the grains' growth rate and
the planetesimals' retention efficiency.  In view of the large
uncertainties in these processes, we adopt the radial gradient of
conventional MMSN model, $q_d = 3/2$.  Different radial gradient
between $\Sigma_d$ and $\Sigma_g$ could be produced by inward
migration of dust grains due to gas drag, which tends to make inner
disk more metal-rich while outer disk metal-poor \citep{Stepinski97,
Kornet}.

In Papers I-III, $f_d$ was assumed to be constant with time until core
mass reaches isolation mass.  In the present paper, we take into
account planetesimal clearing due to the cores' accretion.  Since we
also include type-I migration of the cores, $f_d$ at a given location
(semimajor axis $a$) continuously decreases with time as planetesimals
are accreted by cores which in term undergo orbital decay.
Note that in the present paper, $a$ is identified as $r$, since
we neglect evolution of orbital eccentricities.

The full width of feeding zone of a core is given by 
\citep{KI98,KI02}
\begin{equation}
\begin{array}{lll}
\Delta a_{\rm c} & = & 10 r_{\rm H} \\
  & = & 10 \left( \frac{2M_{\rm c}}{M_{\ast}} \right)^{1/3}a, 
\end{array}
\end{equation}
where $r_{\rm H}$ is two-body Hill radius.  Following Papers I-III, we
take into account of the expansion of feeding zones due to collisions
among the isolated cores after significant depletion of disk gas
\citep{Kominami02,ZLS07}, although this effect influences
the results only slightly (it is important in inner regions
in which cores are significantly depleted by type-I migration). 

An increase in the cores' mass is uniformly subtracted from the mass
in its feeding zone, keeping $q_d$ locally.  When the cores migrate
out from the feeding zone, the reduction of $f_d$ at the initial
location of the cores is stopped.  But, the cores can continue to
accrete planetesimals along their migration path.  In any given system,
when a core reaches 0.03AU (occasionally, it can attain sufficient
mass to accrete gas and become a gas giant during the course of its
migration), a next-generation core is launched at the pre-migration
radius with 100 smaller mass.  (Since core growth is faster during
earlier stage of disk evolution (eq.~[\ref{eq:m_grow0}]), the choice
of initial mass of the next-generation core does not affect the total
core accretion timescale.)  Thereafter the growth of the core and the
depletion of the planetesimals in its feeding zone resume.  

This process repeats until the residual planetesimals are depleted.
Along the cores' migrating paths, they accrete residual planetesimals
(until they reach $a < a_{\rm dep,mig}$ given by
eq.~[\ref{eq:a_depletion}]; see below)
but cannot reduce $f_d$ as fast as the cores formed {\it in
situ}. Therefore we neglect the evolution of $f_d$ due to the
planetesimal accretion by the migrating cores.  The clearing of the
residual planetesimal disk by migrating gas giants, is also neglected
because it rarely occurs and is unlikely to affect the overall
distribution of the emerging terrestrial planets (see \S 3.2.2).  The
overall evolution of $\Sigma_d$ due to core accretion and migration is
described in \S 3.1.

\subsection{Core growth and type-I migration}

We briefly recapitulate a comprehensive analysis on the growth of
planetesimals and cores.  Readers can find a systematic derivation of
these results and the appropriate references from Paper I.  During
their growth through cohesive collisions, cores' mass accretion rate
(of planetesimals) at any location $a$ and time $t$ is described by
(Paper I)
\begin{equation}
\begin{array}{l}
dM_{\rm c}/dt = M_{\rm c}/\tau_{\rm c,acc}; \\
\tau_{\rm c,acc} = 2.2 \times 10^5 \eta_{\rm ice}^{-1}
f_d^{-1} f_{\rm g}^{-2/5} 10^{(2/5)(3/2-q_g)}
\left( \frac{a}{1{\rm AU}} \right)^{27/10 + (q_d-3/2) + (2/5)(q_g-3/2)}
\left(\frac{M_\ast}{M_{\odot}} \right)^{-1/6}
{\rm yrs},
\end{array}
\label{eq:m_grow0}
\end{equation}
where $f_d$ and $f_g$ change with time.  In the derivation of the
above expression, we have adopted the mass of the typical field
planetesimals to be $m=10^{20}$g. The numerical factor $10^{(2/5)
(3/2-q_g)}$ comes from the fact that we scale surface densities at
10AU in eq.~(\ref{eq:sigma_gas}).  

Based on the results of previous numerical simulations (see references
in Paper I), we assume that runaway growth phase is quickly
transferred to oligarchic growth phase \citep{KI98}.  During the
oligarchic growth phase, the cores' accretion timescale is dominated
by their late-stage growth ($\tau_{\rm c,acc} \propto M_{\rm
c}^{1/3}$) so that dependence on their initial mass is negligible.  In
eq.~(\ref{eq:m_grow0}), we adopt $M_{\rm c}(0) = m=10^{20}$g.  When
$f_g < 10^{-3}$, we use gas-free accretion rate rather than that in
Eq.~(\ref{eq:m_grow0}) (Paper I),
\begin{equation}
\tau_{\rm c,acc} = 2 \times 10^7 \eta_{\rm ice}^{-1}
f_d^{-1} \left( \frac{a}{1{\rm AU}} \right)^{3 + (q_d-3/2)}
\left(\frac{M_\ast}{M_{\odot}} \right)^{-1/2}
{\rm yrs}.
\label{eq:m_grow_e}
\end{equation}

In Papers I-III, we artificially terminate the cores' accretion at
$M_{\rm c}= M_{\rm c,iso}$, where the isolation mass is given by
\begin{equation}
M_{\rm c,iso} \simeq
0.16 \eta_{\rm ice}^{3/2} f_d^{3/2} 
\left(\frac{a}{1\mbox{AU}}\right)^{3/4-(3/2)(q_d-3/2)} 
\left(\frac{M_\ast}{M_{\odot}} \right)^{-1/2} M_{\oplus}.
\label{eq:m_iso0}
\end{equation}
In the present paper, we do not need to adopt an artificial
termination to the core growth because when $M_{\rm c}$ approaches
isolation $M_{\rm c,iso}$, {\rm in situ} core accretion is
automatically slowed down by the reduction of $f_d$. However, in outer
regions, ejection of planetesimals by large protoplanets limits their
accretion at $M_{\rm e,sca}$ (scattering limit; Paper I). We take this
effect into account by adopting an artificial termination for the
cores' accretion of planetesimals when they attain a mass $M_{\rm
e,sca} \simeq 1.4 \times 10^3 (a/1{\rm AU})^{-3/2}(M_{\ast}/M_{\odot})
M_{\oplus}$.

Imbalance in the tidal torques from outer and inner disks causes type-I
migration of a core.  Through 3D linear calculation, \citet{Tanaka02}
derived the time scale of type-I migration,
\begin{equation} 
\begin{array}{ll}
\tau_{\rm mig1} & = \frac{a}{\dot{a}} 
= \frac{1}{C_1}
  \frac{1}{2.728 + 1.082 q_g}
  \left(\frac{c_s}{a \Omega_{\rm K}}\right)^{2} 
  \frac{M_*}{M_p}
  \frac{M_*}{a^2 \Sigma_g}
  \Omega_{\rm K}^{-1} \\
 & \simeq 5 \times 10^4 \times 10^{(3/2-q_g)} \frac{1}{C_1 f_g} 
  \left(\frac{M_{\rm c}}{M_{\oplus}} \right)^{-1} 
  \left(\frac{a}{1{\rm AU}}\right)^{q_g} 
  \left(\frac{M_*}{M_{\odot}}\right)^{3/2}
  \;{\rm yrs}.  
\end{array}
\label{eq:tau_mig1} 
\end{equation} 
where we used eq.~(\ref{eq:temp_dist}) and $q_g = 1$--1.5.  

In eq.~(\ref{eq:tau_mig1}), we introduce an scaling factor $C_1$ to
allow for the retardation of type-I migration due to non-linear
effects. In the expression of \citet{Tanaka02}, $C_1 = 1$.  Many
simulations have been carried out \citep[see e.g.,][]{Papaloizou06}
with different numerical methods and resolutions, protoplanetary
potentials and orbits, and disk structures.  They produced a
considerable range of the timescale of type-I migration ($\tau_{\rm
mig 1}$).  Retardation processes for this type-I migration include
variation in the surface density and temperature gradient
\citep{Masset06b}, intrinsic turbulence in the disk \citep{Laughlin04,
Nelson04}, self-induced unstable flow \citep{Koller04, Li05}, and
non-linear radiative and hydrodynamic feedbacks \citep{Masset06a}.
Under some circumstances, $C_1$ is reduced to be $\la 0.1$
(Dobbs-Dixon et al. 2007, in preparation).  In light of these
uncertainties, we adopt $C_1$ as a parameter and mainly examine the
slower migration cases with $C_1 < 1$.

Type-I migration can occur before core mass reaches $M_{\rm c,iso}$.
In this case, $\Sigma_d$ (equivalently, $f_d$) decreases but does not
completely vanish.  From the decreased $\Sigma_d$, new cores accrete
with reduced growth rates and isolation masses.  The cycle of core
growth, type-I migration, and reduction of $\Sigma_d$ continues until
$\Sigma_d$ decreases to the levels from which only cores small enough
not to migrate \citep{Kominami06}.  We present a detailed
discussion on the self-regulation mechanism in later sections.

Under some conditions, the migrating cores can capture and accumulate
planetesimals along their paths \citep{Ward_Hahn, TI99}.  N-body
simulation by \citet{Kominami06}, however, showed that the trapping of
planetesimals by the cores is tentative and it does not
significantly reduce their accretion rates.  In our simulations, we
use the accretion rate for migrating cores that is the same as the
rate for non-migrating cores (eq.~[\ref{eq:m_grow0}]).  For cores in
the systems with $q_g = 1$ and $q_d = 1.5$,
\begin{equation}
\begin{array}{l}
dM_{\rm c}/dt = M_{\rm c}/\tau_{\rm c,acc} \; ; \\
\tau_{\rm c,acc} = 3.5 \times 10^5 \eta_{\rm ice}^{-1} 
f_d^{-1} f_{g}^{-2/5} 
\left( \frac{a}{1{\rm AU}} \right)^{5/2}
\left(\frac{M_\ast}{M_{\odot}} \right)^{-1/6} 
{\rm yrs}, 
\end{array}
\label{eq:tau_c_acc2} 
\end{equation}
\begin{equation}
\begin{array}{l}
da/dt = a/\tau_{\rm mig1}\; ; \\
      \tau_{\rm mig1} \simeq 1.6 \times 10^5 C_1^{-1} f_g^{-1} 
  \left(\frac{M_{\rm c}}{M_{\oplus}} \right)^{-1} 
  \left(\frac{a}{1{\rm AU}}\right)
  \left(\frac{M_*}{M_{\odot}}\right)^{3/2}
  \;{\rm yrs}.  
\end{array}
\label{eq:tau_mig12} 
\end{equation}

The magnitude of $f_d$ is consistently decreased by the increase in
$M_{\rm c}$ (see \S\ref{subsec:sigma_evol}).  Since the accretion rate
is determined by the instantaneous local value of $f_d$, it limits the
mass of the migrating cores in a gaseous medium.  Prior to the severe
depletion of the disk gas, we quench the cores' accretion of
planetesimals at $a < a_{\rm dep,mig}$ (eq.~[\ref{eq:a_depletion}])
on the basis that there is an inadequate supply of residual
planetesimals in these locations to significantly add to their
masses.  But, during the gas depletion, we assume $f_{g}$ decays
exponentially. After disk gas is significantly depleted ($f_g  \la
10^{-3}$), Eq.~(\ref{eq:m_grow_e}) is used for $\tau_{\rm c,acc}$.  We
set also $da/dt$ is set to be zero at disk inner edge ($\sim 0.03$AU).

\subsection{Formation of gas giant planets}

Prescriptions for formation of gas giant planets are the same as those
used in Paper I-III, although the cores' rate of planetesimal
accretion is revised from Paper I-III by the incorporation of type-I
migration and planetesimal depletion. After the formation of these gas
giants, we assume all residual planetesimals in the gap is cleared as
consequence of dynamical instabilities. These planetesimals are either
accreted by the gas giants or scattered elsewhere \citep{Zhou07}. We
neglect the emergence of second-generation cores close to the orbit of
gas giants.
 
In principle, cores with mass much less than that of the Earth can
accrete gas.  But, unless the heat released during gas and
planetesimal accretion is diffused and radiated away, quasi thermal
and hydrodynamic equilibrium would be established to prevent further
flow onto the cores. Around low-mass cores, the temperature and
density of the envelope are low so that heat cannot be easily
redistributed through their envelopes.  But as the cores grow through
planetesimal bombardment beyond a mass
\begin{equation}
M_{\rm c,hydro} \simeq
10 \left( \frac{\dot{M}_{\rm c}}{10^{-6}M_{\oplus}/ {\rm yr}}\right)^{0.25}
M_{\oplus},
\label{eq:crit_core_mass}
\end{equation}
both the radiative and convective transport of heat become efficient
to allow their envelope to contract dynamically \citep{Ikoma00}.  In
the above equation, we neglected the dependence on the opacity in the
envelope (see Paper I).  In regions where the cores have already
attained isolation, their planetesimal-accretion rate $\dot M_{\rm c}$
is much diminished \citep{Zhou07} and $M_{\rm c,hydro}$ can be
comparable to an Earth mass $M_\oplus$.  But, gas accretion also
releases energy and its rate is still regulated by the efficiency of
radiative transfer in the envelope such that
\begin{equation}
\frac{dM_{\rm p}}{dt} \simeq \frac{M_{\rm p}}{\tau_{\rm KH} },
\label{eq:mgsdot}
\end{equation}
where $M_{\rm p}$ is the planet mass including gas envelope.  In Paper
I, we approximated the Kelvin-Helmholtz contraction timescale
$\tau_{\rm KH}$ of the envelope with
\begin{equation}
\tau_{\rm KH} \simeq 10^{k1} 
\left(\frac{M_{\rm p}}{M_{\oplus}}\right)^{-k2} \; {\rm yrs}.
\label{eq:tau_KH}
\end{equation}
In order to take into account the uncertainties associated with
planetesimal bombardment, dust sedimentation and opacity in the
envelope, we adopt a range of values $k1 = 8$--10 and $k2 = 3$--4 (see
Papers I and II).  Here we use $k2 = 3$ and treat $k1$ as a parameter.

Gas accretion onto the core is quenched when the disk is depleted
either locally or globally.  We assume that gas accretion is
terminated if either the thermal condition or global depletion
condition is satisfied.  A (partial) gap is formed when the rate of
tidally induced angular momentum exchange by the planet with the disk
exceeds that of the disk's intrinsic viscous transport \citep{LP85}
(the viscous condition),
\begin{equation} 
M_p > M_{\rm g,vis} 
\simeq 30 \left(\frac{\alpha}{10^{-3}}\right) 
\left(\frac{a}{1{\rm AU}}\right)^{1/2} 
\left(\frac{L_\ast}{L_{\odot}}\right)^{1/4} M_{\oplus}.
\label{eq:m_gas_vis} 
\end{equation} 
Planets with $M_p > M_{\rm g,vis}$ have sufficient mass to induce the
partial clearing of the disk near their orbit.  The tidal torque on
either side of the gap becomes sufficiently strong to induce the
planets to adjust their positions within the gap.  This feedback
process leads to a transition from type-I to type-II migration.
Therefore, we adopt the viscous condition for the onset of type-II
migration.  In Papers I-III, we adopted $\alpha = 10^{-4}$ as a
nominal value. In order to match the simulated $M_p-a$ distribution to
the observed data, eq.~(\ref{eq:m_gas_vis}) was arbitrarily multiplied
by a scaling factor $A_\nu = 10$.

The previously adopted value of $\alpha$ is smaller than that inferred
from the models of protostellar disk evolution \citep{Hartmann98}.  In
this paper, we use $\alpha = 10^{-3}$ without imposing the scaling
factor.  As a result, the condition for the onset of type-II migration
is the same as that in Papers I-III.

A clear gap would be formed and gas accretion would be terminated when
the planet's Hill radius becomes larger than disk scale height
\citep{LP85} (the thermal condition) and it is given by (Paper I)
\begin{equation}
M_p > M_{\rm g,th}
\simeq 0.95 \times 10^3 \left(\frac{a}{1{\rm AU}}\right)^{3/4}
\left(\frac{L_\ast}{L_\odot}\right)^{3/8}
\left(\frac{M_\ast}{M_\odot}\right)^{-1/2} M_\oplus.
\label{eq:m_gas_th}
\end{equation}
While the thermal condition $r_{\rm H} > 1.5 H$ was used in Papers I-III,
we here used the condition $r_{\rm H} > 2 H$, in order to
make clear that the asymptotic mass of gas giants is determined
by global depletion of disk gas rather than the local thermal condition
in the cases with inclusion of type-I migration 
in which gas giants are formed in relatively late stages.
(It can also reflect the effect of gas flow into the gap mentioned below.)

Some numerical simulations indicate that a residual amount of gas may
continue to flow into the gap after both the viscous and thermal
conditions are satisfied\citep{D'Angelo03,Kley06,Tanigawa07}.
However, recent numerical simulations \citep{Dobbs-Dixon07} also show
that the azimuthal accretion flow from the corotation regions onto the
planet is effectively quenched despite a diminishing flux of gas into
the gap.  Since the thermal condition usually requires larger $M_p$
than the viscous condition ($M_{\rm g,th} > M_{\rm g,vis}$), we adopt
it as the criterion for the termination of gas accretion in the
determination of the asymptotic mass of gas giant planets.  But, we
take into a residual amount of gas which may leak through the gap and
provide an effective avenue for angular transfer between the inner and
outer regions of the disk via the gas giant's corotation resonance 
(\S 2.4).

Gas accretion may also be limited by the diminishing amount of
residual gas in the entire disk even for planets with 
$M_{\rm p} < M_{\rm p,th}$.  In Papers I-III,
we assumed that the maximum available mass is determined 
simply by $M_{\rm g,no iso} \sim \pi a^2 \Sigma_g$.
However, because the global limit plays a more important role
when type-I migration is incorporated, we use a more appropriate
condition $M_{\rm g,no iso} \sim \int_0^{2a} 2\pi a \Sigma_g da$
(We neglect disk gas inflow from far outer regions).
For $q_g = 1$,
\begin{equation}
M_{\rm g,no iso} \simeq  3.5 \times 10^2 f_{\rm g,0} 
\exp \left(-\frac{t}{\tau_{\rm dep}}\right) 
\left(\frac{a}{1\mbox{AU}}\right)^{1/2} M_{\oplus}.
\label{eq:m_gas_non_iso}
\end{equation}
When $M_{\rm g,noiso}$ diminishes below $M_{\rm p}$ or when $M_{\rm
p}$ exceeds $M_{\rm g,th}$, gas accretion is terminated.  

\subsection{Type-II migration}

In our previous analysis in Papers I-III, we adopted a simply analytic
prescription for type-II migration: (i) before disk gas mass decays to
the value comparable to the planet mass, the planet migrates with
(unperturbed) disk accretion and (ii) when disk gas mass is comparable
to the planet mass, a fraction ($C_2 \sim 0.1$) of the total (viscous
plus advective) angular momentum flux transported by the disk gas
(which is assumed to be independent of the disk radius) is utilized by
the planet in its orbital evolution. 

If the planet's tidal torque can severely clear a gap in the vicinity
of its orbit, $\Sigma_g$ in inner disk region would decrease faster
than that in the outer region.  The full torque asymmetry leads to
$C_2 \simeq 1$ in case (ii).  But, when the truncation condition is
marginally satisfied, the disk interior to the planet's orbit have a
total mass $\ga M_p$ such that it can effectively replenish the
angular momentum lost by planet to the outer disk region.  There may
also be a leakage of gas through the gap region \citep{D'Angelo03}
which would suppress the degree of torque asymmetry.  The
protoplanet's corotation resonance may drive an effective angular
momentum transfer across the two disk regions separated by the
protoplanet.  

Protoplanets are formed in disk regions where the midplane is inactive
to magneto-rotational instabilities.  It is possible that the modest
accretion rate onto the host stars flow through this region via an
active layer which is exposed to external ionizing photons and cosmic
ray particles \citep{Gammie96}.  Different values of effective $\alpha$'s
may contribute to the mass flow through the disk and the planet-disk
interaction, especially if there is some leakage across the gap
region.  All of these possible scenarios can lead to $C_2 \ll 1$.

The uninterrupted replenishment of gas into the corotation region may
also maintain a finite vortensity gradient and an unsaturated
corotation torque \citep{Masset06a} which may reduce the efficiency of
type-II migration from disk gas accretion even in case (i)
\citep{Crida07}.  However, the results of another set of 2D numerical
hydrodynamic simulations \citep{D'Angelo06} essentially reproduces the
the 1D simulation \citep{LP86} in which the contribution from the
corotation resonance is neglected.

In order to take into account of these uncertainties, we reduce here
the migration rate by a factor $C_2$ for cases (ii) as in Papers I-III,
\begin{equation}
\begin{array}{ll}
\tau_{\rm mig2,ii} 
 & {\displaystyle = \frac{1}{C_2} 
\frac{(1/2)M_p \Omega_{\rm K}(a) a^2}
{3 \pi \Sigma_{g}(r_m) r_m^2 \nu_m \Omega_{\rm K}(r_m)}} \\
 & {\displaystyle \simeq 5 \times 10^5 f_g^{-1} 
\left(\frac{C_2 \alpha}{10^{-4}} \right)^{-1} 
\left(\frac{M_{\rm p}}{M_{\rm J}} \right) 
\left(\frac{a}{1{\rm AU}}\right)^{1/2} 
\left(\frac{M_*}{M_{\odot}}\right)^{-1/2}}
\;{\rm yrs}.  
\end{array}
\label{eq:tau_mig2b} 
\end{equation} 
When $\tau_{\rm mig2,ii}$ is shorter than the migration timescale for (i), 
we use the latter timescale, 
\begin{equation}
\begin{array}{ll}
\tau_{\rm mig2,i} 
 & {\displaystyle  = \mid \frac{a}{\dot{a}} \mid 
= \frac{a}{(3/2) \nu / a} }\\
 & {\displaystyle \simeq 0.7 \times 10^5
\left(\frac{\alpha}{10^{-3}}\right)^{-1}
\left(\frac{a}{1{\rm AU}}\right)
\left(\frac{M_*}{M_{\odot}}\right)^{-1/2}}
\;{\rm yrs}.
\end{array}
\label{eq:tau_mig2a} 
\end{equation}
Due to more accurate estimations,
the numerical factors in $\tau_{\rm mig2,i}$ and
$\tau_{\rm mig2,ii}$ slightly differ from those
in Papers I-III for the same values of $C_2$ and $\alpha$.  
However, there are still uncertainties in the formula
for $\tau_{\rm mig2,ii}$.  Furthermore, 
it is not easy to theoretically evaluate the
value of $C_2$ as well as $\alpha$.  
Hence, we vary the
magnitude of $C_2$ and compare the simulated results with the observed
data to obtain a calibration. 
For $\alpha \sim 10^{-3}$, in the limit that 
$C_2 \sim 1$, most of gas giants are removed from the regions beyond
1AU, which is inconsistent with observed data of extrasolar planets.
As shown in Paper II, in order to reproduce $M_p$--$a$ distribution of
observed extrasolar planets, disk depletion timescale $\tau_{\rm dep}$
must be $\sim \tau_{\rm mig2}$ at a few AU.
In the present paper, we mostly adopt 
$C_2 \alpha = 10^{-4}$.  
For $\alpha \sim 10^{-3}$, it corresponds to $C_2 \sim 1/10$.
As a result, $\tau_{\rm mig2,ii}$ we adopt here is almost
identical to that used in Papers I-III.

\section{Effects of type-I migration}

In our previous simulations in Papers I-III, the effect of type-II
migration was included but that of type-I migration was neglected. In
this section, we investigate the effects of type-I migration on the
planets' $M_p$--$a$ distribution.

\subsection{Surface density evolution due to type-I migration}
\label{subsec:sigma_evol}

In order to analyze the asymptotic $M_p-a$ distribution from the Monte
Carlo simulations, we first present the results on the $\Sigma_d$
reduction due to type-I migration.  The relevant time scale here is a
function of $C_1, f_g$ and $M_p$ (eq.~[\ref{eq:tau_mig1}]). The
magnitude of the cores' $M_p$ is a function of $f_d$, $f_g$ and $t$.

Since type-I and II migrations involve the tidal interaction of
embedded cores with the disk gas, the dynamical clearing of the
residual planetesimals is suppressed after the gas is severely
depleted (except for outer regions in which ejection by massive
embryos can be efficient).  Since $t \sim \tau_{\rm dep}$ is a
critical stage for the build-up and retention of the cores and the
onset of gas accretion onto the cores, we are particularly concerned
with the residual $\Sigma_d$--distribution at this stage.  
The asymptotic $\Sigma_d$--$a$
distribution is determined by $C_1, f_{g,0}$ and $\tau_{\rm dep}$.
In this subsection, we show the simulation results with $f_{g,0} = f_{d,0}$,
$q_d = 1.5$ and $q_g = 1.0$ at $t = 0$.  
The results with other reasonable values of $f_{d,0}/f_{g,0}$, $q_d$
and $q_g$ are qualitatively similar.  With these initial conditions,
we can compute the emergence of cores during the epoch of gas
depletion.

Figure \ref{fig:sigma_type1_mig1_fg3}a shows $\Sigma_d$ at $t = 10^5,
10^6$, and $10^7$ years (dashed, dotted, and solid lines) with $C_1 =
1$ and $f_{g,0} = 3$.  These results correspond to surviving
protoplanets at $t \sim \tau_{\rm dep}$ for $\tau_{\rm dep} = 10^5,
10^6$, and $10^7$ years, although depletion of $f_g$ on time scales
$\tau_{\rm dep}$ is not taken into account here (in the Monte Carlo
simulations in section 3.2, the exponential decay,
eq.~(\ref{eq:gas_exp_decay}), is considered).  We generated 1000 semi
major axes with a log-uniform distribution in the range of 0.05--50 AU
and simulated the growth and orbital migration of cores there.  
In these simulations, gas accretion onto cores is neglected
in order to clearly see the effects of type-I migration.
The dynamical interactions among the cores is also neglected.  
With N-body simulations, \citet{Kominami05} showed that the
dynamical interactions with other cores and planetesimals do not
change the cores' type-I migration speed.  The results presented in
this panel confirm the finding of N-body simulation by
\citet{Kominami06}. They also show that $\Sigma_d$ is depleted in an
inside-out manner.  The clearing of planetesimals, along the paths of
previous generations of cores, limits the growth of the migrating
cores and the gravitational interactions between them.  In order to
take into account the effect of dynamical clearing in the Monte Carlo
simulations (to be presented in next subsection), we terminate the
growth of cores after their semimajor axes have decreased below
$a_{\rm dep,mig}$ given by eq.~(\ref{eq:a_depletion}).

Figure \ref{fig:sigma_type1_mig1_fg3}d shows that the total mass of
the remaining population of cores is $\sim 0.1M_{\oplus}$.  For these
retained cores, $\tau_{\rm mig1} > 3 \tau_{\rm c,acc}$ 
($\tau_{\rm c,acc} = \dot{M}_{\rm c}/M_{\rm c}$), 
where a factor 3 reflects an actual
timescale to reach $M_{\rm c}$ because $\tau_{\rm c,acc} \propto
M_{\rm c}^{1/3}$.  From eqs.~(\ref{eq:tau_c_acc2}) and
(\ref{eq:tau_mig12}), the maximum mass of remaining cores ($M_{\rm
c,max}$) is given by
\begin{equation}
M_{\rm c,max} \simeq
0.21 C_1^{-3/4} 
\left(\frac{f_{g,0}}{3}\right)^{3/10}
\left(\frac{\eta_{\rm ice} f_{d,0}}{f_{g,0}}\right)^{3/4}
\left( \frac{a}{1{\rm AU}} \right)^{-9/8}
\left(\frac{M_\ast}{M_{\odot}} \right)^{5/4} M_{\oplus}.
\label{eq:m_c_max}
\end{equation}
The above expression with $f_{g,0}=f_{d,0}$ reproduces the result at
$\ga a_{\rm dep,mig} \sim 1$AU in Figure~
\ref{fig:sigma_type1_mig1_fg3}d as well as the weak dependence of
$M_{\rm c,max}$ on $f_{g,0}$ and $C_1$
(Figures~\ref{fig:sigma_type1_mig1_fg30}d and \ref
{fig:sigma_type1_mig01_fg3}d).  In order to further examine the
$f_{g,0}$ and $C_1$ dependences, we carry out a set of models with
$f_{g,0} = 30$ ($C_1 = 1$) and $C_1 = 0.1$ ($f_{g,0} = 3$) in Figures
\ref{fig:sigma_type1_mig1_fg30} and \ref{fig:sigma_type1_mig01_fg3}.
In the former case, the disks is marginally self-gravitating and
contain a significant fraction of the central stars' mass.  The
emergence of relatively massive cores is more sensitively determined
by $C_1$, the inefficiency of type-I migration.  A large amount of
solid materials in the disk does not efficiently promote the formation
of massive cores, because the associated dense gas increases type-I
migration speed, as already pointed out by \citet{Kominami06}.

In the inner regions at $a \la 0.3$AU, the cores' accretion proceeds
on very short time scales and they reach their isolation mass near
their birth place (Figure \ref{fig:sigma_type1_mig1_fg3}c).
Thereafter, most of these cores migrate into their host stars within
$10^5$ years.  Consequently, the local $\Sigma_d$ is essentially
depleted by the formation and migration of the very first generation
cores (panel c).  This domain is determined by
the condition $M_{\rm c,max} \ga M_{\rm c,iso}$.  From
eqs.~(\ref{eq:m_iso0}) and (\ref{eq:m_c_max}), this condition implies
\begin{equation}
a \la a_{\rm iso,mig} \simeq 0.45 C_1^{-2/5} 
\left(\frac{f_{g,0}}{3}\right)^{-2/3}
\left(\frac{\eta_{\rm ice} f_{d,0}}{f_{g,0}}\right)^{-2/5}
\left(\frac{M_\ast}{M_{\odot}} \right)^{14/15}{\rm AU}.
\label{eq:a_1st_gene}
\end{equation}
This expression approximately reproduces the result in
Figure~\ref{fig:sigma_type1_mig1_fg3} and it accounts for the
dependence on $C_1$ and $f_{g,0}$ (see
Figures~\ref{fig:sigma_type1_mig1_fg30}c and
\ref{fig:sigma_type1_mig01_fg3}c).  The location of $a_{\rm iso,mig}$
does not vary with the slope of the surface density distribution
because both competing processes are determined by local properties of
the disk.

In the limit of efficient type-I migration (with $C_1=1$), most of the
cores undergo orbital decay before they attain their isolation mass in
the intermediate regions (for Figures~\ref{fig:sigma_type1_mig1_fg3},
0.3AU $\la a \la 10$AU).  Consequently a significant fraction of
$\Sigma_d$ remains to promote the formation of subsequent-generation
cores.  The results of our simulations (Figure
\ref{fig:sigma_type1_mig1_fg3}c) show that many generations of cores
may emerge at the same disk location.  This repeated formation and
self-destruction process is even more efficient in massive disks where
$f_{g,0} \gg 1$ (see Figure \ref{fig:sigma_type1_mig1_fg30}c).  

In the inner region, surface density is significantly depleted by the
N generations of core formation and disruption where
\begin{equation}
N_{\rm gene,1} \simeq \frac{M_{\rm c,iso}}{M_{\rm c,max}}
\simeq 4.1 C_1^{3/4} 
\left(\frac{f_{g,0}}{3}\right)^{6/5}
\left(\frac{\eta_{\rm ice} f_{d,0}}{f_{g,0}}\right)^{3/4}
\left(\frac{a}{1{\rm AU}}\right)^{15/8}
\left(\frac{M_\ast}{M_{\odot}} \right)^{-7/4}.
\label{eq:N_gene1}
\end{equation}
As the planetesimal building blocks become depleted, the cores formed
at later epochs have masses smaller than $M_{\rm c,max}$.  Therefore
the above expression for $N_{\rm gene,1}$ slightly under estimates the
number of populations of cores which may emerge.  

At large disk radii, the number of generation is limited by disk
depletion time.  In this region,
\begin{equation}
N_{\rm gene,2} = \frac{t}{\tau_{\rm c,mig}(M_{\rm c,max})}
\simeq 3.9 C_1^{1/4} 
\left(\frac{f_{g,0}}{3}\right)^{13/10}
\left(\frac{\eta_{\rm ice} f_{d,0}}{f_{g,0}}\right)^{3/4}
\left(\frac{a}{1{\rm AU}}\right)^{-17/8}
\left(\frac{t}{10^6{\rm years}}\right)
\left(\frac{M_\ast}{M_{\odot}} \right)^{-1/4}.
\label{eq:N_gene2}
\end{equation}
This estimate completely reproduces the results in Figures~
\ref{fig:sigma_type1_mig1_fg3}c, \ref{fig:sigma_type1_mig1_fg30}c, and
\ref{fig:sigma_type1_mig01_fg3}c.  The actual number of generation is
given by min$(N_{\rm gene,1}, N_{\rm gene,2})$.  Significant depletion
of the original inventory of heavy elements occurs in the limit
$N_{\rm gene,1} \la N_{\rm gene,2}$, that is, within the location
\begin{equation}
a \la a_{\rm dep,mig} 
\simeq C_1^{-1/8} 
\left(\frac{f_{g,0}}{3}\right)^{1/40}
\left(\frac{t}{10^6{\rm years}}\right)^{1/4}
\left(\frac{M_\ast}{M_{\odot}} \right)^{3/8}{\rm AU}.
\label{eq:a_depletion}
\end{equation}
This boundary of disruption zone is in excellent agreement with the
critical location within which $\Sigma_d$ (equivalently, from $f_{d}$)
has reduced from its initial values by an order of magnitude.  Note
that the dependences of $a_{\rm dep,mig}$ on $C_1$ and $f_{g,0}$ are
very weak.  As long as $N_{\rm gene,2} > 1$, the disruption zone is
confined to $a \sim (t/10^6{\rm years})^{1/4}$ AU, independent of disk
surface density and migration speed $C_1$ (in the limit of small
$C_1$, $N_{\rm gene,2} < 1$ and $N_{\rm gene,1} < 1$,
so that no depletion occurs).

These results indicate that, during the early epoch of disk evolution,
cores form and migrate repeatedly to clear out the residual
planetesimals. This self-regulated process provides an avenue for the
host stars to acquire most of the heavy elements retained by the
planetesimals. In large, well-mixed molecular clouds where star
clusters form, this self-regulated mechanism would lead to stellar
metallicity homogeneity (\S 4.1). The inner disk region contains cores
which have started their type-I migration but not yet reached to the
disks' inner edge. The average value of $\Sigma_d$ at these locations,
including the contribution of these migrating cores, is two orders of
magnitude smaller than its initial value.

Eventually, the disk gas is so severely depleted that relatively
massive cores no longer undergo significant amount of type-I
migration. As any given disk radius, the condition for retaining 90\%
of the initial solid surface density is $N_{\rm gene,2} \simeq
0.1N_{\rm gene,1}$.  We find, from eqs.~(\ref{eq:N_gene1}),
(\ref{eq:N_gene2}), and (\ref{eq:a_depletion}), that this condition is
satisfied in regions with
\begin{equation}
a \ga a_{\rm surv,mig} 
= 10^{1/4} a_{\rm dep,mig} \simeq 2 a_{\rm dep,mig}.
\label{eq:a_survival}
\end{equation}
Type-I migration leads to a transition in the $\Sigma_d$ distribution
at this orbital radius.

Inside the ice line, type-I migration limits the mass of individual
surviving cores.  But these cores can coalesce through giant impacts
during and after the severe depletion of the disk gas
\citep{Kominami02, IL04a}.  Provided the total mass of residual
planetesimals and cores is $\sim O(1)M_{\oplus}$ at 1AU $\la a \la$ a
few AU, Earth-mass terrestrial planets may form near 1AU.  Previous
simulations \citep{Chambers98, Agnor99, Kominami02, Raymond04} show
that the most massive terrestrial planets tend to form in inner
regions of the computational domain where the isolated cores are
initially placed. In these simulations, strong gravitational
scattering process can inject planetesimals close to the host stars to
form smaller planets with relatively close-in orbits.

In general, type-I migration leads to clearing of planetesimals close
to their host stars and sets the inner edge of the cores' population
at $\sim 1$AU.  The lack of planets inside the Mercury's orbit in our
Solar system might also be attributed to this result
\citep{Kominami06}.  In addition, the self-regulated clearing process
also leads to $f_d \sim O(1)$ near 1AU for wide variety of initial
conditions ($f_{d,0}$).  This residual distribution of heavy elements at
$\tau_{\rm dep}$ ensures the formation of Mars to Earth-sized
terrestrial planets in habitable zones (see \S 3.2).  Even in the
limit that the disks' initial $\Sigma_d$ distribution is much larger
than that of MMSN, the reduction of $\Sigma_d$ at $\la$ a few AU
inhibits {\it in situ} formation of gas giants interior to the ice
line.  In contrast to the results in Papers I-III, the inclusion of
type-I migration suppresses the rapid and prolific formation of hot
Jupiter which in term facilitates the formation and retention
probability of habitable terrestrial planets.

We also simulated several models with $q_d = 2$ which correspond to a
steeply declining initial surface density distribution. Such a steep
gradient of $\Sigma_d$ could be produced by the inward migration of dust
due to the hydrodynamic drag on them by the disk gas 
\citep{Stepinski97, Kornet}.  In this model, a significant amount of
solid mass is contained in inner disk regions where cores quickly
form and undergo type-I migration.  Nearly all the initial mass of
solid components in disks is accreted by the host stars through the
self-regulation by type-I migration.  The results of this model is
consistent with the discovery of metallicity homogeneity among the
stars in young open clusters (see \S 4.1).

The above discussions indicate that the formation of Mars to
Earth-sized habitable planets depends only weakly on type-I migration
speed.  However, it is critical for the formation of cores of gas
giant planets, because $M_{\rm c,max} \propto C_1^{-3/4}$
(eq.~[\ref{eq:m_c_max}]).  For giant planets to actually form,
sufficiently massive cores must be able to accrete gas on time scales
at least shorter than a few folding time of $\tau_{\rm dep}$.  For
$\tau_{\rm dep} \sim 10^6$--$10^7$ years, we deduce, from
eq.~(\ref{eq:mgsdot}) with $k1 =9$ and $k2=3$, that gas giant
formation is possible only for $M_{\rm c} \ga$ a few $M_{\oplus}$.
But, type-I migration suppresses the emergence of such massive cores
in disk regions with relatively large $\Sigma_g$.  The results with
$C_1=1$ in Figures \ref{fig:sigma_type1_mig1_fg3}d and
\ref{fig:sigma_type1_mig1_fg30}d show that the asymptotic core masses
are generally much smaller than that needed to launch efficient gas
accretion even though there is little decline in the magnitude of
$\Sigma_d$.

In the case of $C_1=0.1$ (Figure \ref{fig:sigma_type1_mig01_fg3}d),
the cores formed at a few AU originally have $\sim M_{\oplus}$.  
Provided the planetesimals along their migration path
are not captured onto their mean motion resonances, these cores can
grow up to a few $M_{\oplus}$ through accretion during migration.  The
core mass is marginal for rapid gas accretion. As shown in
eq.~(\ref{eq:m_c_max}), in metal-rich disk regions (where $f_{d,0}/
f_{g,0} > 1$), more massive cores can be retained with the same amount
of solid materials.  In this expression, the effects of gas depletion
has not been taken into account. But, this process is included in the
numerical simulations in \S3.2 and it also enhances the mass of
retainable cores, especially those which emerge during the advanced
stages of gas depletion.  These results suggest that gas giants can be
formed for $C_1 \la 0.1$.  (\citet{Alibert05} derived a similar
condition for formation of Jupiter and Saturn.)  In the next
subsection, we will show that $C_1 \la 0.1$ reproduces mass and semi
major axis distribution of extrasolar gas giants comparable to those
observed.
 
\subsection{Mass - semimajor axis distributions}

\subsubsection{Dependence on type-I migration speed}

We now consider the signature of type-I migration on the $M_p-a$
distribution for a population of emerging planets.  In the Monte Carlo
simulations, we first generate a 1,000 set of disks with various
$f_{d,0}$ (the initial value of $f_d$) and $\tau_{\rm dep}$.  We adopt
the same prescriptions for the distributions of $f_{d,0}$ and
$f_{g,0}$ as those in Paper II.  For the gaseous component, we assume
$f_{g,0}$ has a log normal distribution which is centered on the value
of $f_{g,0} = 1$ with a dispersion of 1 ($\delta \log_{10} f_{g,0} =
1.0$) and upper cut-off at $f_{g,0} = 30$, independent of the stellar
metallicity.  For the heavy elements, we choose $f_{d,0}= 10^{{\rm
[Fe/H]}_d} f_{g,0}$, where [Fe/H]$_d$ is metallicity of the disk.  We
assume these disks have the same metallicity as their host stars.

Following our previous papers, we also assume $\tau_{\rm dep}$ has
uniform log distributions in the ranges of $10^6$--$10^7$ yrs.  For
each disk, 15 $a$'s of the protoplanetary seeds are selected from a
uniform log distribution in the ranges of 0.05--$50$AU, assuming that
averaged orbital separation between planets is 0.2 in log scale (the
averaged ratio of semimajor axes of adjacent planets is $\simeq 1.6$).
This procedure is the same as that adopted in Paper II.  Constant
spacing in the log corresponds to the spacing between the cores is
proportional to $a$, which is the simplest choice and a natural
outcome of dynamical isolation at the end of the oligarchic growth.
The log constant spacing for planets and cores with similar masses
also maximizes dynamical stability.  In the present paper, we neglect
dynamical interaction between planets (this issue will be addressed in
future papers) and the growth of individual planets are integrated
independently.  Although the choice of averaged orbital separation is
arbitrarily, it would not change overall results with regard to the
effects of type-I migration.

In all the simulations presented here, $\alpha = 10^{-3}$ 
and $M_{\ast} = 1M_{\odot}$ are assumed.
Since the on-going radial velocity surveys are focusing
on relatively metal-rich stars, we present the results with [Fe/H] =
0.1 in most cases.  The dependence on [Fe/H] is shown in
Figures~\ref{fig:metal_dep}.  In order to directly compare with
observations, we determine the fraction of stars with currently
detectable planets as $\eta_J$.  In the determination of $\eta_J$, we
assume the detection limit is set by the magnitude of radial velocity
($v_r > 10$m/s) and orbital periods ($T_K < 4$ years).
According to the following uncertainty, we exclude close-in planets 
with $a < 0.05$AU in the evaluation of $\eta_J$.

We artificially terminate type-I and II migration near disk inner edge
at a 2 day period ($\sim 0.03$AU for $M_{\ast} = 1 M_{\odot}$) 
in a similar way to Papers I-III.  
We have not specified a survival criterion for the
close-in planets because we do not have adequate knowledge about
planets' migration and their interaction with their host stars near
inner edge of their nascent disks.  Hence, we record all the planets
which have migrated to the vicinity of their host stars.  In reality,
a large fraction of the giant planets migrated to small disk radii may
either be consumed \citep[e.g.,][]{Sandquist98} or tidally disrupted
\citep[e.g.,][]{Trilling98, Gu03} by their host stars.  Cores that
have migrated to inner edge may also coagulate and form super-earths
\citep[e.g.,][]{Terquem07}.  We also neglect such core coagulation
near inner edge.

In a set of fiducial models, we adopt $M_{\ast} = 1 M_{\odot}$, [Fe/H]
= 0.1 and $(k1,k2)=(9,3)$ for the gas giants' growth rate
(eq.~[\ref{eq:tau_KH}]).  The analytic deductions in the previous
subsection suggest that relatively massive cores can be retained to
form gas giants provided $C_1 \la 0.1$.  Figures \ref{fig:ma_C1} show
the predicted $M_p$--$a$ distributions for $C_1 = 0$ (panel b), 
$C_1 = 0.01$ (panel c), $C_1 = 0.03$ (panel d), $C_1 = 0.1$ (panel e), 
and $C_1 = 0.3$ (panel f).  In order to directly compare
the theoretical predictions 
with the observed data, we plot (in panel a) $M_p$ which is a factor
of 1.27 times the values of $M_p \sin i$ determined from radial
velocity measurements (http://exoplanet.eu/).  
This correction factor corresponds to the average
value $1/ \langle \sin i \rangle = 4/\pi$ for a sample of planetary
systems with randomly oriented orbital planes.  To compare with 
the theoretical results with $M_{\ast} = 1M_{\odot}$, 
we plot only the data of planets around stars 
with $M_{\ast} = 0.8$--$1.2 M_{\odot}$ observed by radial velocity surveys.

All results show "planet desert," 
which is a lack of intermediate-mass ($M_p \sim 10$--$100M_{\odot}$)
planets at $\la$ a few AU.
However, formation probability of gas giants dramatically
changes with $C_1$. The fraction of stars with gas giants 
$\eta_J$ changes from $22.9\%$ for the model with $C_1 = 0$ 
to $0.2\%$ for $C_1 = 0.3$ (see Table \ref{tab:1}).  
In the observed data, $\eta_J \sim 5$--8$\%$ around 
[Fe/H]$\sim 0.1$ (Figure~\ref{fig:metal_dep}).

In contrast, the distributions of retained gas giants are similar 
to each other except for relatively large $C_1$.
The theoretical predictions are also consistent with the
observed distribution in panel a.  We carry out a KS test for
statistical similarity between the simulated models and the observed
data for the parameter domain of 0.1AU $< a < 2.5$ AU and 
$M_{\rm p} > 100M_{\oplus}$.  
This range corresponds to a maximum rectangular
region in which planets are detectable by radial surveys with
precision $v_r > 10$m/s and duration $<4$ years. Since we have not
imposed any criterion for determining the survival probability of
short-period planets, the predicted population of planets with 
$a < 0.1$AU are excluded in the quantitative statistical significance test.
Except for the model with $C_1=0.3$ (panel f),
these models are statistically
similar to the observed data within 
a significant level $Q_{\rm KS} \ga 0.3$ for both semimajor axis
and mass cumulative distribution functions.  
In particular, the model with $C_1=0.03$ (panel d) shows
an excellent agreement with $Q_{\rm KS} = 0.86$ for 
the mass function.  

In models with $C_1=0.3$, only the low-mass cores can survive type-I
migration.  The envelope contraction time scales for these low-mass
cores are generally much longer (eq.~[\ref{eq:tau_KH}]) than the gas
depletion time scales.  Consequently, $\eta_J$ is very small ($<1\%$) 
for the simulated model with $C_1 = 0.3$. Since type-II
migration occurs after planets have acquired a mass which is adequate
to open up gaps, the close-in planets with $\ga 100 M_{\oplus}$ are
rare for models with $C_1=0.3$, in contrast to the models that type-I
migration is neglected or sufficiently reduced (panels b to e). 

Assuming the survival fraction of close-in planets is independent
their $M_p$, the magnitude of $C_1$ can be calibrated from the observed
mass distribution close-in planets.  In Figure~\ref{fig:close-in}a, we
plot the mass function for all the planets which are halted
artificially at $a=0.03$ AU.  In this panel, we neglect any further
evolution including both disruption and collisions.  We also consider
an alternative limit that after the gas depletion, the
multiple-generation of cores which migrated to the proximity of any
given star are able to merge into a single terrestrial planet with a
mass $M_{\rm mer}$ (Fig.~\ref{fig:close-in}b).  These results suggest
the potential findings of many hot earths, including Neptune-mass
planets, with either transit or radial velocity surveys.

The above results clearly highlights the competing effect of type-I
migration and gas accretion.  In these models, we approximate the gas
accretion process with $(k1, k2) = (9,3)$.  The early models of
proto-gas-giant planet formation \citep{Pollack96} yield slower growth
rates and they are better fitted with $(k1, k2) = (10,3.5)$.  But,
recent revisions \citep{Ikoma00, Hubickyj05} of these models indicate
that the protogas giants' growth rates can be significantly enhanced
by the opacity reduction associated with grain growth or boundary
conditions at different regions of the disk \citep{Ikoma01}.  The
Kelvin-Helmholtz contraction timescale may also be reduced by
turbulent heat transport in the outer envelope of the protoplanets.
In view of these uncertainties, we also simulated models with $k1 = 8$
and 10, with $k2 = 3$ for all cases.  The predicted $\eta_J$ are
listed in Table \ref{tab:1}.  For models with $C_1 = 0.03$--0.1 in which
type-I migration marginally suppresses the formation of gas giants,
the magnitude of $\eta_J$ depends sensitively on the minimum mass for
the onset of dynamical gas accretion (which is represented by $k1$).
The results in Table \ref{tab:1} indicate that a smaller value of $k1
(=8)$ can lead to a significant increase in $\eta_J$ because smaller
mass cores can initiate the runaway gas accretion within $\tau_{\rm
dep} = 10^6$--$10^7$ years (eq.~[\ref{eq:tau_KH}]).

\subsubsection{Preservation of terrestrial planets}

Although type-I migration 
is an effective mechanism to suppress the emergence of
gas giant planets, earth-mass planets can form at $\sim 1$AU even in
the limit of $C_1 \sim 1$.  (But most terrestrial planets with $a \la
0.3$AU would be eliminated by a full-strength type-I migration.)  
Not all of these planets formed interior
to the ice line may be retained.  A fraction of these planets may be
trapped onto the mean motion resonance of migrating gas giant planets
and be forced to migrate with them \citep{Zhou05}. In the 
relatively unlikely event of inefficient eccentricity 
damping and very fast type-II migration, 
some of the embryos along the path of the migrating gas
giants may also be scattered to large radial distances and form later
generation terrestrial planets \citep{Raymond06}.

The survival of terrestrial planets depends on their post-formation
encounter probability with migrating giant planets.  In the absence of any
type-I migration, this probability is modest. But the inclusion of a
small amount of type-I migration significantly reduce the fraction of
stars with massive close-in gas giants because the retention of the
progenitor cores becomes possible only at the late stages of disk
evolution when the magnitude of $\Sigma_g$ is reduced.  With a limited
supply of the residual disk gas, the growth of gas giants and their
type-II migration are suppressed.  

We find that repeated migration of gas giants is less common in models
with $C_1 \ga 0.01$ than those with $C_1 = 0$. The low type-II
migration probability reduced the need for efficient disruption of
largely accumulated close-in planets (see Paper II). It also ensures
that most of the terrestrial planets formed in the habitable zones are
not removed by the migrating gas giants.  Note that type-I migration
also inhibits {\it in situ} formation of gas giants near 1AU (see \S
3.1).  Thus, a small amount of type-I migration facilitates formation
and retention of terrestrial planets in habitable zones in extrasolar
planetary systems, rather than inhibits them.

\subsubsection{Metallicity dependence}

We also study the dependence on other parameters, metallicity
([Fe/H]).  As fiducial models, we set $C_1=0.03$ and $k1 =9$ to compare
the metallicity dependence.  The most massive cores which can form and
be retained prior to gas depletion have masses $M_{\rm c,max}$ given
by eq.~(\ref{eq:m_c_max}).  Smaller $f_{g,0}$, larger $f_{d,0}$, or
smaller $C_1$ increases $M_{\rm c,max}$ and would enhance formation of
gas giants.  The models in Figures \ref{fig:ma_C1} have already
illustrated the dependence on $C_1$.

In Paper II, we showed that relatively large values of [Fe/H] (or
equivalent $f_{d,0}/f_{g,0}$) enhance the growth rates and increase
the isolation masses for the cores, for the disks with the same $f_{g,0}$.  
Fast emergence of massive cores
can lead to the rapid onset of gas accretion.  Based on that model, we
found $\eta_J$ increases with [Fe/H], and the simulated of $\eta_J$
dependence on the [Fe/H] is qualitatively consistent with observed
data \citep{Fischer05}.

Here we study the effect of type-I migration on the $\eta_J$--[Fe/H]
correlation.  In Figures~\ref{fig:ma_C1}, [Fe/H] = 0.1 is assumed.  We
carried out similar simulations with various [Fe/H] and the simulated
$\eta_J$--[Fe/H] relation is plotted in Figure~{\ref{fig:metal_dep}).
A comparison between the results here and those in Paper II shows that
type-I migration enhances the $\eta_J$--[Fe/H] correlation. In
metal-poor disks, $f_{d,0}/f_{g,0} < 1$ and $\eta_J$ is significantly
reduced by type-I migration.  But in more metal-rich disks, relatively
massive cores can be retained before the disk gas is severely
depleted.  The resultant steep dependence is in a better agreement
with the observed data (open circles in the plot) than in the
dependence without type-I migration (filled circles with dashed line),
although different choice of assumed averaged orbital separation 
may change $\eta_J$ slightly.

Formed planetary systems are affected by stellar mass $M_*$ 
as well as metallicity. 
The dependence on $M_*$ was studied in Paper III,
using a simple model without the effects of type-I migration.
In the paper, we predicted that gas giants are much more rare 
around M type dwarfs than around FGK dwarfs while
super-Earths are abundant around M type dwarfs, 
which is consistent with radial velocity survey 
and microlensing survey \citep{Beaulieu06}.
Adding the type-I migration to the simple prescription in Paper III, 
we found that the above conclusions do not change.
Around M type dwarfs, super-earths at 1--3AU, 
which are inferred to be abundant by microlensing survey, 
survive type-I migration.
In a separate paper, we will address the details of the $M_*$ dependence
of planetary systems,
taking into account the $M_*$ dependences of many physical quantities.

\section{Summary and Discussions}

In our previous Monte Carlo simulations of planet formation processes
(Papers I, II, and III), we neglected the effects of type-I migration.
In the present paper, we have investigated its effects on formation of
terrestrial planets and cores of gas giants.  We found that type-I
migration provides a self-clearing mechanism for planetesimals in the
terrestrial-planet region. Although the planetesimal disk at $a \la
1$AU is significantly cleared, the total mass of residual planetary
embryos at regions within a few AU is comparable to the Earth's mass, 
almost independent of the disk and migration parameters. Earth-like planets
can be assembled in habitable zones after the depletion of the disk gas.  

But this self-regulated clearing process does prevent giant planets
from forming at $\sim 1$AU even in very massive disks. In general, the
clearing of cores leads to the late formation of gas giants.  When the
surface density of the disk gas is reduced below that of the MMNM,
type-I migration would no longer be able to remove cores which are
sufficiently massive to initiate the onset of rapid gas
accretion. This late-formation tendency also reduces the fraction of
gas giant which undergo extensive type-II migration. Since migrating
gas giants capture and clear cores along their migration paths, type-I
migration also facilitates the retention of Earth-mass planets in the
habitable zones.

In the limit that type-I migration operates with an efficiency
comparable to that deduced from the traditional linear torque analysis
({\it i.e.} with $C_1 = 1$), all cores would be cleared prior to the
gas depletion such that gas giant formation would be effectively
suppressed.  However, the catastrophic peril of type-I migration would
be limited by at least a ten-fold reduction in the type-I migration 
speed ($C_1 \la 0.1$).  In this limit, a substantial fraction of the cores may
survive and gas giants would form and be retained with an efficiency
($\eta_J$) comparable to that observed.

The above discussions indicate that the formation probability of gas
giants is delicately balanced by various competing processes. Since
these processes have comparable efficiency, the fraction of solar type
stars with gas giants appears to be ``threshold'' quantity. Small
variations in the strength of one or more of these effects can
strongly modify the detection probability of extrasolar planets.  For
example, $\eta_J$ is observed to be a rapidly increasing function of
their host stars' metallicity \citep{Fischer05}.  Although we were able
to reproduce this observed trend in Paper II, the simulations
presented here indicate that this effect is enhanced by a small amount
of type-I migration because it is more effective in metal-poor disks.

The $M_p$ - $a$ distribution in Figures~\ref{fig:ma_C1} also show its
sensitive dependence on various other model parameters.  The most
noticeable dependence is the relative frequency between gas and ice
giant planets at several AU from their host stars.  These
distributions also predict a modest population of close-in cores, a
fraction of which may survive and be observable as hot earths.

\subsection{Metallicity homogeneity of open clusters}

In modern paradigm of star formation \citep{Shu87}, most of the stellar
content is processed through protostellar disks.  Gas diffusion in
these disks is regulated by the process of turbulent transport of
angular momentum whereas the flow of heavy elements is determined by
orbital evolution of their main carriers, {\it i.e.} through gas drag
of grains and tidal interaction of planetesimals and cores with the
gas. In general the accretion rates of these two components are
not expected to match with each other.  In fact, the formation of gas
giant planets around $\sim$10\% of the solar type stars and the common
existence of debris disks requires the retention of heavy elements
with masses at least comparable to that of the MMSN.  Yet, stars in
young stellar clusters are chemically homogenous \citep{Wilden02,
Quillen02, Shen05}.  The observationally determined upper limit in the
metallicity dispersion ($< 0.03$--0.04 dex) among the stars in the
Pleiades and IC4665 open clusters implies a total residual heavy
element mass (including the planets) to be less than twice that in
Solar system planets.

Clues on the resolution of this paradox can be found in the
protostellar disks.  Recent models of the observed millimeter
continuum SED's suggest that some fraction of classical T Tauri disks
has dust mass $\ga$ 10 times the total heavy-element mass in Solar
system \citep{Hartmann06}.  Their host stars would acquire significant
metallicity dispersion if a large fraction of the heavy elements in
these disks is retained as terrestrial planets and cores while most of
their gas components is accreted by the stars.  Hence, an efficient
process to clear heavy elements is required.

Although, due to gas drag, dust grains can undergo inward migration
and be accreted onto their host stars, we here assume that the
formation of planetesimals is a more efficient self-clearing process.
In active protostellar disks where $\Sigma_g$ and $\Sigma_d$ are much
larger than those of MMSN, planetesimals quickly grow into cores which
undergo rapid type-I migration. Through a series of numerical
simulations, we show that most cores formed interior to $a_{\rm
dep,mig} \sim 1$ AU are lost before the disk gas is severely depleted.
On a larger spatial and time scales, this process is also effective in
transporting most of the original heavy elements to their host stars.
Provided their progenitor molecule clouds is thoroughly mixed, this
self-regulated clearing process would ensure stars in young clusters
acquire nearly uniform metallicity.

\subsection{Diversity of giant planets around solar-type stars}

Among the stars with known gas giants, a large fraction of them show
signs of additional planets. This special multiplicity function is
associated with the formation of gaps around the first-born gas
giants.  Beyond the outer edge of the gap, a positive pressure
gradient leads the gas to attain a local super Keplerian velocity.
Dust particles and planetesimals accumulate in this region and grow
into sufficiently massive cores to initiate the formation of
additional planets\citep{Bryden00}.  It is also possible that the
planet formation is a threshold phenomenon, {\it i.e.} the conditions
needed to form a multiple-planet systems are marginally more stringent
than that for the formation of single gas giants.

The simulations presented here consider the formation probability of
individual planets.  Although the impact of type-I migration on the
residual disk and the formation of multiple generation cores have
been taken into account, we have neglected the impact of gas giant
formation on the residual disks.  Nevertheless, this algorithm can be
used to qualitatively describe a threshold scenario for the formation 
of multiple planets.

The total formation timescale of gas giants is $\tau_{\rm form} \sim
\tau_{\rm c,acc} + \tau_{\rm KH}$.  In comparison with both type-II
migration and the disk depletion timescales, we find that, provided
$\tau_{\rm c,acc} < \tau_{\rm mig1} $,
   \begin{itemize}
   \item Solar-system-like giant planets would form,
         if $\tau_{\rm form} \sim \tau_{\rm dep}$, because 
         gas giants would not have
         enough time to undergo extensive type-II migration.
   \item Eccentric giants would form, 
         if $\tau_{\rm form} < \tau_{\rm dep}$ and 
         $\tau_{\rm mig2} > \tau_{\rm form}$.  The second condition implies 
         that nearby second-generation gas giants are likely to emerge, 
         either by spontaneously satisfying the formation conditions or due 
         to an induced formation process\citep{Bryden00}, before they 
         undergo any significant type-II migration.  Orbital instability 
         of closely packed multiple-planet systems would excite their 
         eccentricities\citep{Rasio96, Weidenschilling96, LI97, ZLS07}.
   \item Close-in giants would form if $\tau_{\rm form} < \tau_{\rm dep}$, and
         $\tau_{\rm mig2} < \tau_{\rm form}$. 
         They undergo type-II migration before neighboring gas giants form.
   \end{itemize}
In the second case, if $\tau_{\rm mig2} < \tau_{\rm dep}$ and orbital
instability does not occur on a time scale $\sim \tau_{\rm dep}$, the
multiple planets could be locked into mean-motion resonances during
migration.  According to the above conditions, the regions in which
close-in, eccentric, and solar-system-like giant planets are likely to
form are schematically plotted in the $f_{d,0}$--$a_0$ plane ($a_0$ is
the initial semimajor axis) in Figure~\ref{fig:areas}, 
neglecting type-I migration.
Bright gray, dark gray, and modest gray regions correspond to
close-in, eccentric, and solar-system-like giants regions,
respectively.  A more quantitative set of simulations will be 
presented in a subsequent paper.

\vspace{1em} 
\noindent ACKNOWLEDGMENTS.  We thank for detailed helpful comments by
an anonymous referee.  This work is supported by NASA (NAGS5-11779,
NNG04G-191G, NNG06-GH45G), JPL (1270927), NSF(AST-0507424,
PHY99-0794), and JSPS.

{}

\clearpage
\begin{table}

\begin{tabular}{l|ccccc} \hline
          & $C_1 = 0$ & $C_1 = 0.01$ & $C_1 = 0.03$ & $C_1 = 0.1$ & $C_1 = 0.3$ \\ \hline\hline
$k_2 = 8$ &   23.7    &     19.1     &  15.9        &     6.7     &   2.0  \\
$k_2 = 9$ &   22.9    &     17.2     &  12.1        &     4.3     &   0.2  \\
$k_2 = 10$ &  21.3    &     12.0     &   7.9        &     2.3     &   0.   \\ \hline
\end{tabular}
\caption{
$\eta_J$ (in percent) for various $C_1$ and $k_1$.
Other parameters are fixed: metallicity [Fe/H]$=0.1$,
stellar mass $M_* = 1M_{\odot}$.
}

\label{tab:1}
\end{table}

\clearpage

\begin{figure}
\plotone{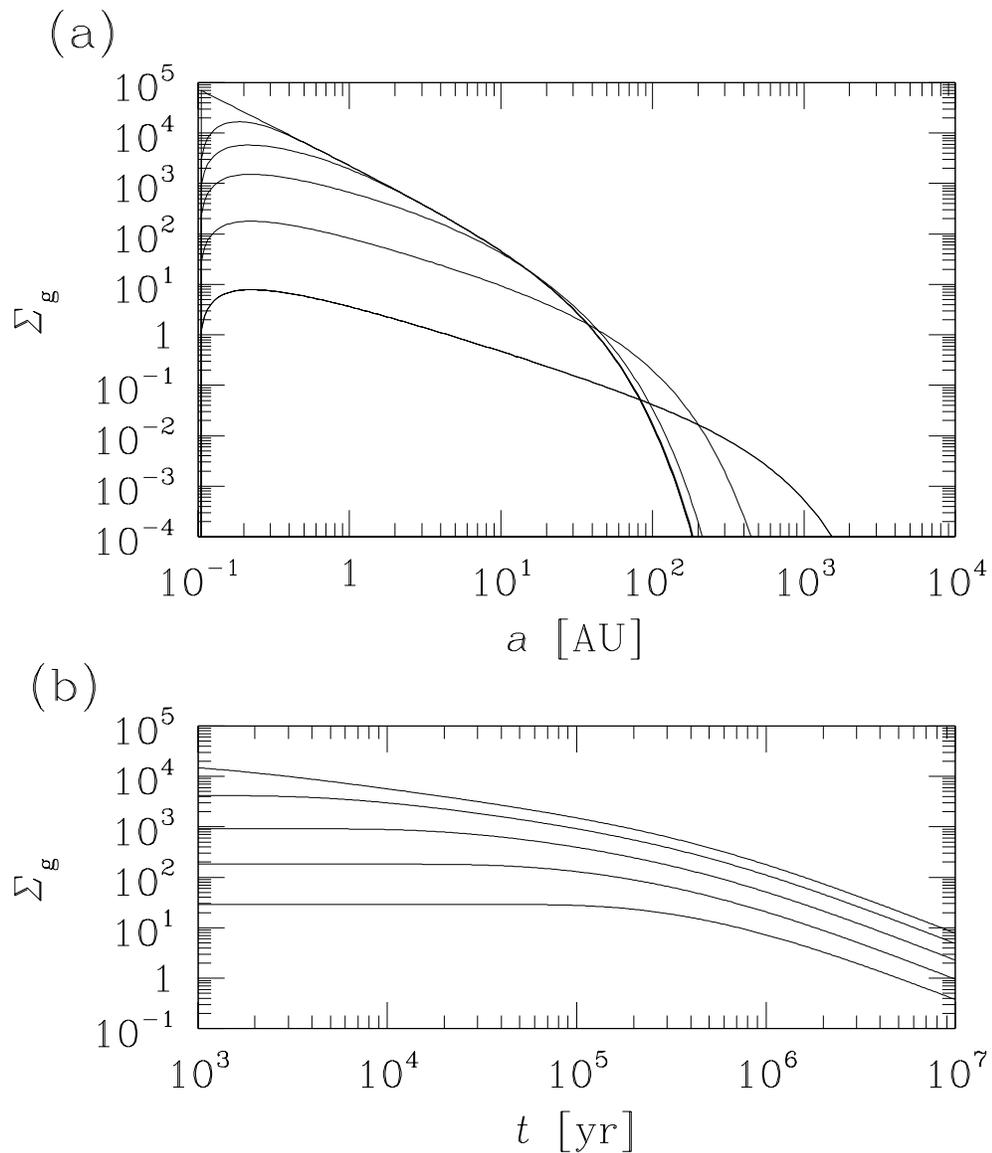}
\caption{
Viscous evolution of a disk with $\alpha = 10^{-3}$.
We set $\Sigma_g = 0$ at $r =0.1$AU and $10^5$AU.
(a) The distributions of $\Sigma_g$ at $t = 0,
10^3, 10^4, 10^5, 10^6,$ and $10^7$ years (from top to bottom).
(b) Time evolution of $\Sigma_g$ at planet-forming
regions $r = 0.25, 0.67, 1.8, 4.7,$ and 12.5 AU
(from top to bottom).
}
\label{fig:disk_vis_evol}
\end{figure}

\begin{figure}
\plotone{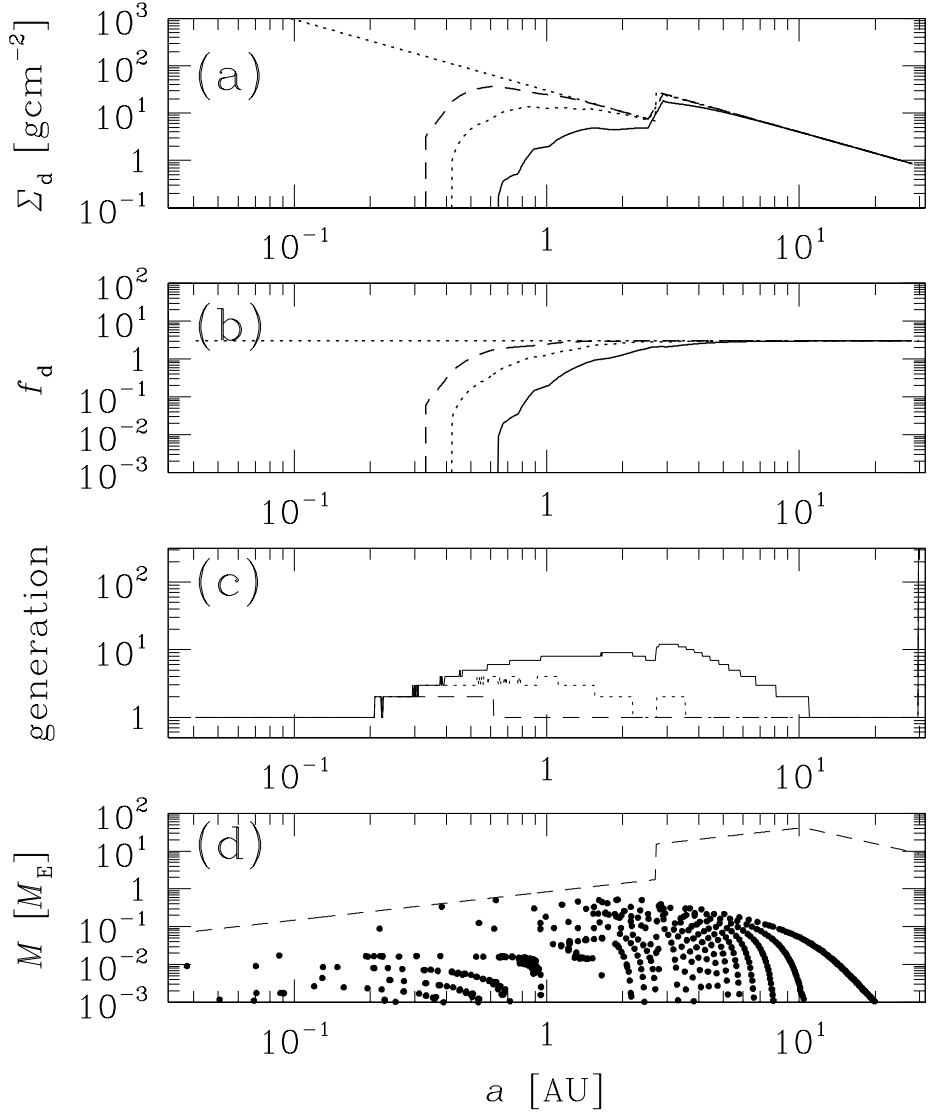}
\caption{ The evolution of $\Sigma_d$ due to dynamical sculpting by
type I migration in the case of $C_1 = 1$ and $f_{g,0} = 3$.  
Here we assume constant $f_g$.
(a) The $\Sigma_d$-distributions at $t = 10^5, 10^6$, and $10^7$ years 
are expressed by
dashed, dotted, and solid lines.  The initial distribution is also
shown by dotted lines.  (b) The evolution of $f_d$-distributions.
(The meanings of the lines are the same as panel a.)
(c) The number of generation of protoplanetary cores formed
at each $a$.  (d) Planet distributions at $t = 10^7$ years.  The dashed line
expresses the core isolation mass and the scattering limit.
 }
\label{fig:sigma_type1_mig1_fg3}
\end{figure}

\begin{figure}
\plotone{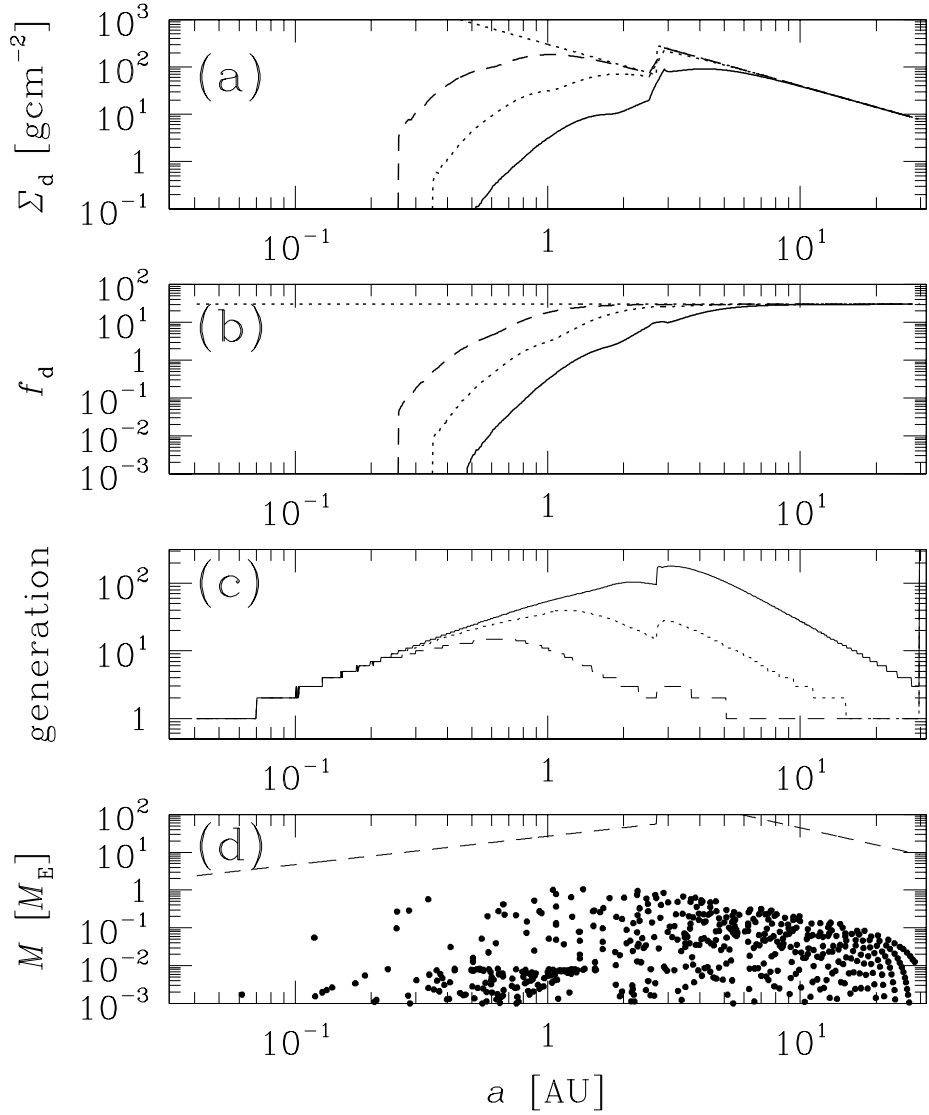}
\caption{ The same as Figure~\ref{fig:sigma_type1_mig1_fg3} except for
$f_{g,0} = 30$ to approximate the early phases of massive disk
evolution.  }
\label{fig:sigma_type1_mig1_fg30}
\end{figure}

\begin{figure}
\plotone{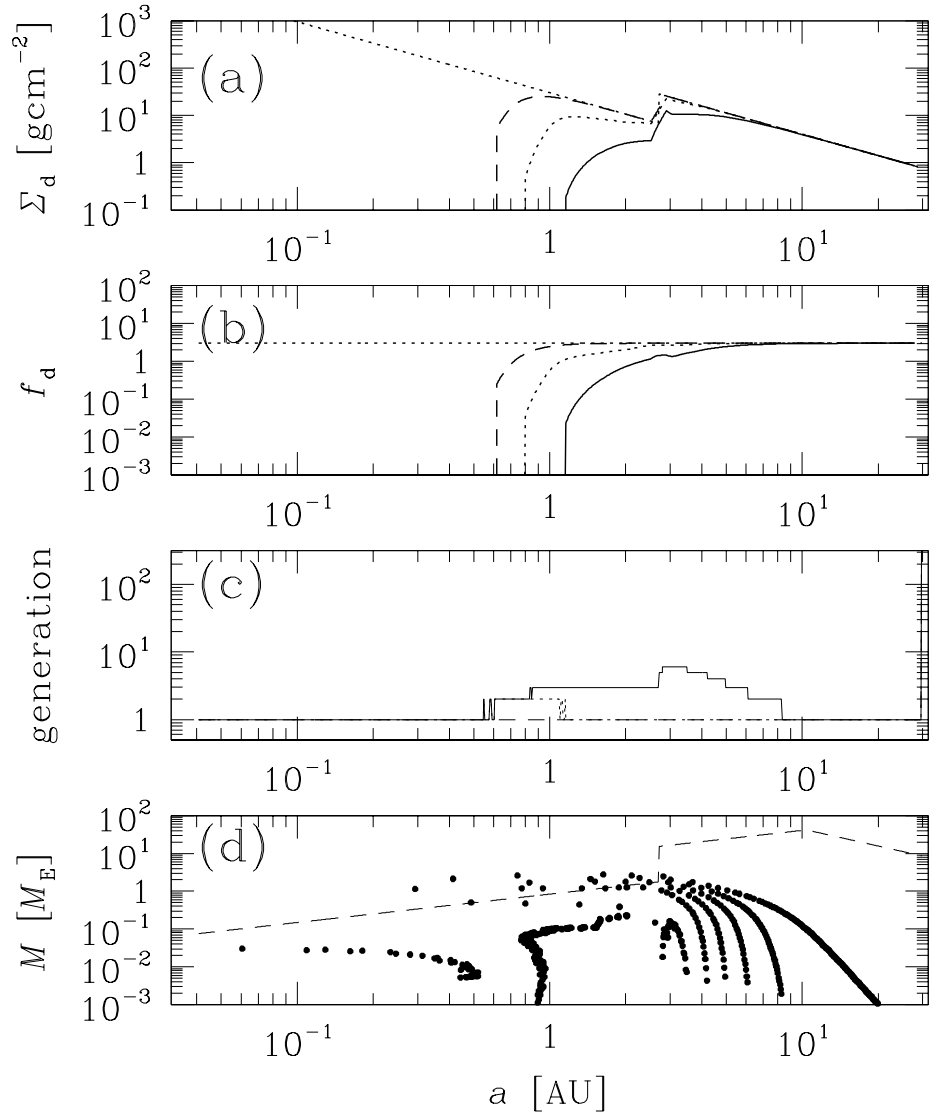}
\caption{ The same as Figure~\ref{fig:sigma_type1_mig1_fg3} except for
$C_1 = 0.1$ to approximate weak type I migration.  }
\label{fig:sigma_type1_mig01_fg3}
\end{figure}

\begin{figure}
\plotone{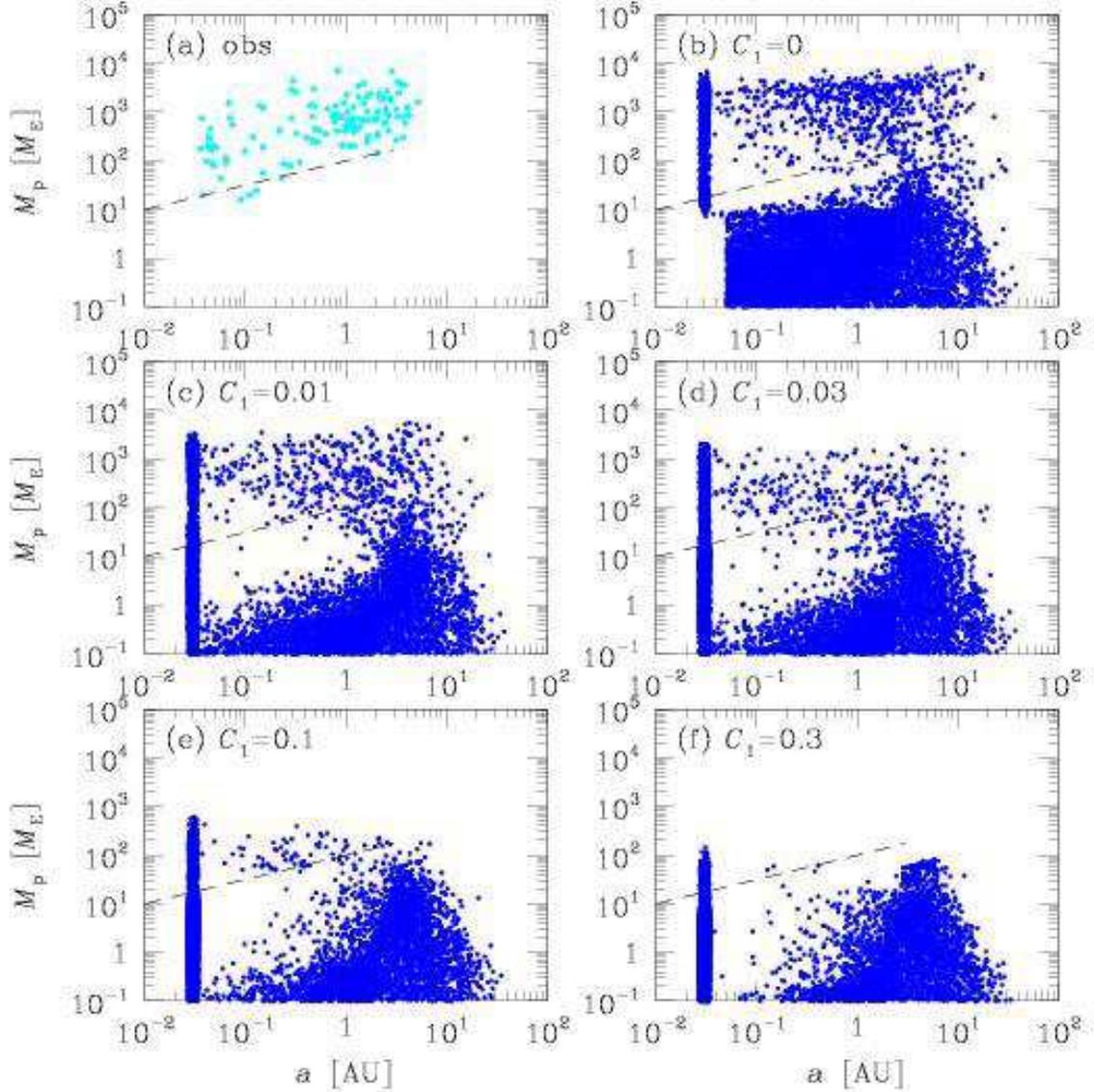}
\caption{ Planetary mass and semi major axis distribution.  
Units of the mass ($M_{\rm p}$) and semimajor axis ($a$)
are earth mass ($M_{\oplus} = M_{\rm E}$) and AU.
(a) Observational data of extrasolar planets
(based on data in http://exoplanet.eu/), around stars with 
$M_{\ast} = 0.8$--$1.2 M_{\odot}$ detected by the
radial velocity surveys.
The determined $M_p \sin i$ is
multiplied by $1/ \langle \sin i \rangle = 4/\pi \simeq 1.27$,
assuming random orientation of planetary orbital planes.
(b) The distribution obtained from 
Monte Carlo simulations without taking into account the effect of 
type-I migration, (c) that includes the type-I migration with
$C_1 = 0.01$, (d) $C_1 = 0.03$, (e) $C_1 = 0.1$, and
(f) $C_1 = 0.3$.  The dashed lines express
observational limit with radial-velocity 
measure precision of $v_r = 10$m/s.  
In these models, 
$M_{\ast} = 1 M_{\odot}$,
the magnitude of the metallicity [Fe/H] = 0.1 and 
the contraction time scale parameters in eq.~[\ref{eq:tau_KH}] are 
assumed to be $(k1,k2)=(9,3)$. }
\label{fig:ma_C1}
\end{figure}

\begin{figure}
\plotone{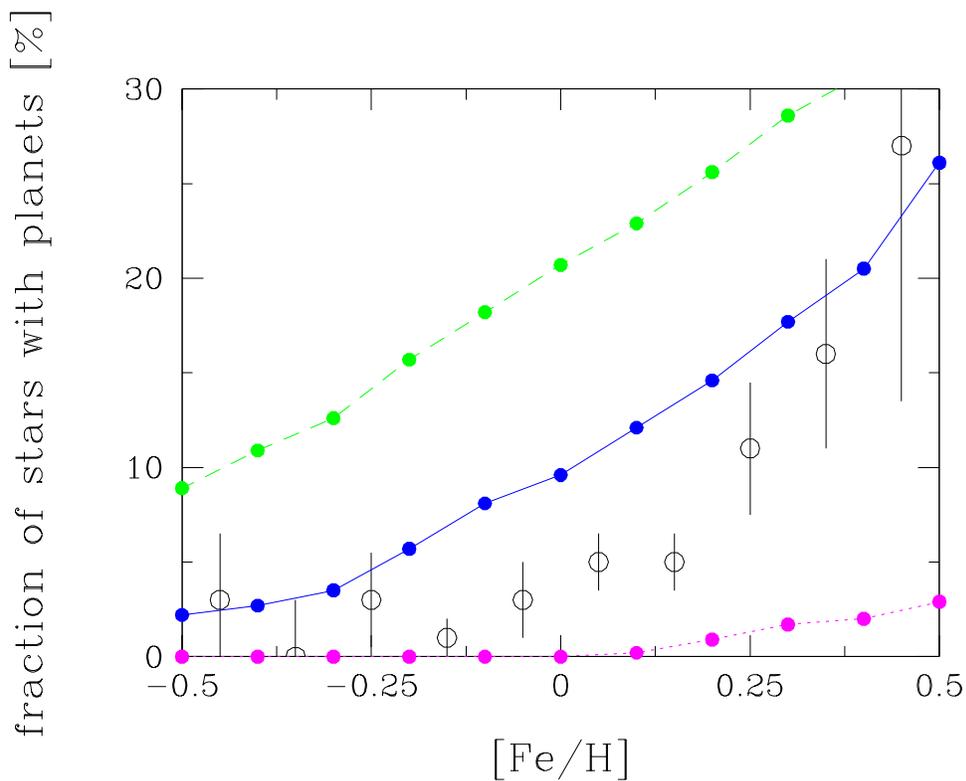}
\caption{ 
The metallicity dependence of fraction of stars
with planets which are detectable by radial velocity surveys
with measurement precision $v_r > 10$m/s and coverage periods $< 4$ years.
Open circles with error bars are observed data \citep{Fischer05}.
Dashed, solid, and dotted lines with filled circles
are the theoretically predicted dependences
with $C_1 = 0, 0.03$ and 0.3, respectively.
Other parameters are
$(k1,k2) = (9,3)$, and $M_{\ast} = 1 M_{\odot}$.
}
\label{fig:metal_dep}
\end{figure}

\begin{figure}
\plotone{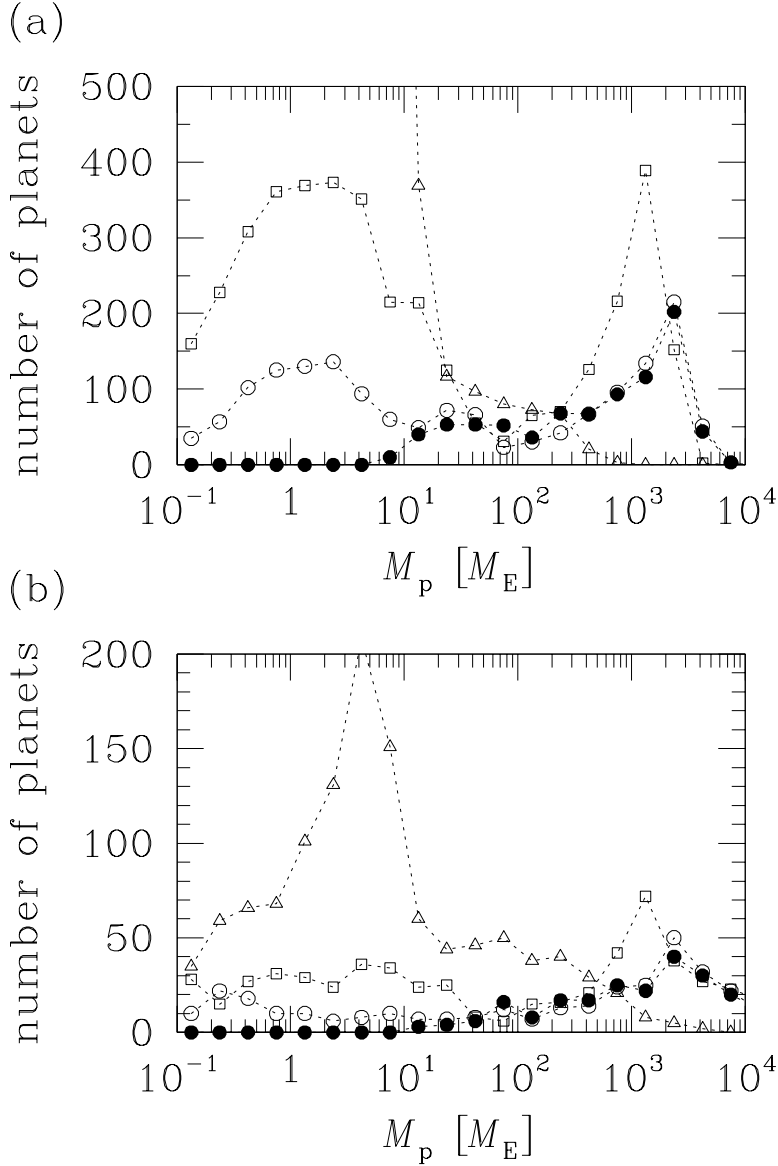}
\caption{ 
The predicted mass function which are halted
artificially at $a=0.03$ AU.
(a) The mass function for all the planets,
neglecting further evolution due to both disruption 
and collisions is neglected.  
Filled circles, open circles, squares, and triangles
represent the results with $C_1 = 0, 0.001, 0.01$, and 0.1,
respectively.
($M_{\ast} = 1 M_{\odot}$, [Fe/H] = 0.1 and $(k1,k2)=(9,3)$.)
(b) That includes the maximum effect of 
coagulation after the gas is depleted. All the planets 
are coagulated into one planet in each system.   
}
\label{fig:close-in}
\end{figure}

\begin{figure}
\plotone{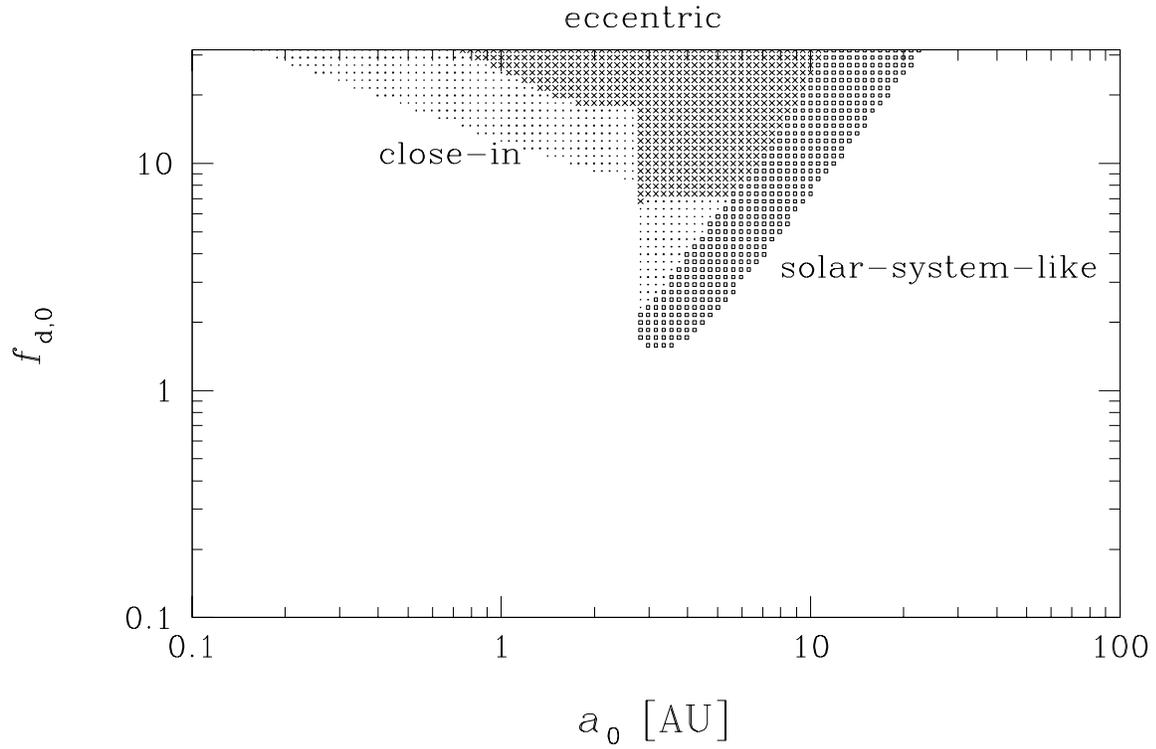}
\caption{The parameter domains (initial disk surface density
$f_{d,0}$ and initial semimajor axis of planets $a_0$), 
in which close-in, eccentric, and
solar-system-like giant planets are likely to form, in the limit of
negligible effects of type-I migration.  Bright gray, dark
gray, and modest gray regions correspond to close-in, eccentric, and
solar-system-like giants regions, respectively.}
\label{fig:areas}
\end{figure}

\end{document}